\documentclass[preprint,prb,superscriptaddress,floatfix]{revtex4-2}
\usepackage{amsmath,amssymb}
\usepackage{graphicx}
\usepackage{bm}
\usepackage[colorlinks=true,linkcolor=blue,citecolor=blue,urlcolor=blue]{hyperref}
\newcommand{\aph}{$\alpha$}
\newcommand{\dph}{$\delta$}
\newcommand{\bph}{$\beta$}
\newcommand{\Cm}{C\,m$^{-2}$}
\newcommand{\Udp}{U_{\rm dp}}
\newcommand{\UKS}{U_{\rm KS}}
\begin{document}

\title{
Poly(vinylidene fluoride) under finite electric displacement fields: \\
a removable $\alpha\!\to\!\delta$ reorientation barrier and a persistent
first-order $\delta\!\to\!\beta$ conformational change
}

\author{Shaohui Qiu}
\email{shaohuiqiu@suu.edu}
\affiliation{Department of Chemistry and Physics, Southern Utah University, Cedar City, Utah 84720, USA}
\author{Jacob C. Dean}
\affiliation{Department of Chemistry and Physics, Southern Utah University, Cedar City, Utah 84720, USA}
\author{Rajagopalan Thiruvengadathan}
\affiliation{Department of Engineering and Technology, Southern Utah University, Cedar City, Utah 84720, USA}
\author{Ramanujam Brahmadesam Thoopul Srinivasa Raghava}
\affiliation{Department of Physics, Amrita School of Physical Sciences, Amrita Vishwa Vidyapeetham, Coimbatore 641112, India}
\affiliation{Functional Materials Laboratory, Amrita School of Engineering, Amrita Vishwa Vidyapeetham, Coimbatore 641112, India}
\author{Yiming Su}
\affiliation{Utah Water Research Laboratory, 1600 Canyon Road, Logan, UT, 84321, USA}
\affiliation{Department of Biological and Chemical Engineering, Utah State University, Logan, UT, 84322, USA}



\begin{abstract}

The $\alpha$, $\delta$, and $\beta$ phases of poly(vinylidene
fluoride) (PVDF) are interconverted by electrical poling and
mechanical drawing, yet the thermodynamics of these conversions under
a controlled electrical boundary condition has not been established
from first principles.
We evaluate the zero-temperature free energy at fixed electric
displacement $D$ from van der Waals density-functional calculations
along two constrained coordinates in a common 24-atom cell.
The first is a tilt of the gauche C--C bonds that maps antipolar
$\alpha$ onto polar $\delta$ and changes the cell by less than $1\%$.
The second is a backbone torsion that straightens
trans--gauche$^+$--trans--gauche$^-$ (TG$^+$TG$^-$) chains into
all-trans (TTTT) $\beta$ and strains the chain repeat by $10.6\%$.
Because 
the $\beta$ phase polarization ($P_\beta$) is $0.55$ of the 
polarization quantum $e/ab$, the
Berry-phase branch is tracked along the path.
As $D$ increases from 0 to $0.30$~\Cm, the lowest homogeneous free 
energy switches from $\alpha$ to $\delta$ at $D_{c1}=0.046$~\Cm\ 
and from $\delta$ to $\beta$ at $D_{c2}=0.157$~\Cm.
At $T=0$ with domain-wall energy neglected, a $\delta$ multidomain 
state lies $14.3$~meV per cell below homogeneous $\alpha$ at 
every $D$. Homogeneous $\alpha$ is never the ground state.
A $\Gamma$-point harmonic estimate suggests that vibrations reverse
this ordering at room temperature.
$\delta$ and $\beta$ coexist for $0.11<D<0.21$~\Cm\ at
$E^*=0.69$~GV\,m$^{-1}$, a field that does not depend on the
screening constant.
The two barriers respond differently to $D$.
The tilt barrier falls from $0.66$~eV per cell at $D=0$ to $17$~meV
at $D=0.30$~\Cm, below $k_BT$ at $300$~K.
The torsion barrier is still $0.66$~eV at $D_{c2}$ and $0.34$~eV at
$D=0.30$~\Cm, so $\delta\to\beta$ remains first order.
Much of this difference traces to the path effective charge, the
dipole created per unit atomic displacement, which is $0.28e$ for the
tilt and $0.18e$ for the torsion.
Scaling the polarization change along each path shows that the tilt
barrier would survive with the effective charge of the torsion.
The smaller effective charge of the torsion accounts for about half
of the barrier it retains.
The landscape accounts for why a field produces $\delta$ and why
$\beta$ is reached by drawing.
It also suggests that raising the path effective charge of the
conformational step could let a field alone produce $\beta$.
\end{abstract}

\maketitle

\section{Introduction}
\label{sec:intro}

Poly(vinylidene fluoride) (PVDF) is a prototypical ferroelectric polymer
\cite{lovinger1983,furukawa1989,kepler1992}.
Its properties depend on which of several polymorphs it adopts, and
the polymorphs differ in both chain conformation and chain packing.
The nonpolar $\alpha$ phase ($P2_1/c$), obtained from the melt,
consists of trans--gauche$^+$--trans--gauche$^-$ (TG$^+$TG$^-$) chains
packed with antiparallel dipoles.
The polar $\delta$ phase ($Pna2_1$) contains the same chains packed
with parallel dipoles along the lattice vector $c$.
The strongly polar $\beta$ phase ($Amm2$) has all-trans (TTTT) chains,
whose dipole per monomer is about $1.6$ times that of the
TG$^+$TG$^-$ conformation and also lies along $c$
\cite{hasegawa1972,lovinger1983}.
Figure~\ref{fig:struct} shows the three phases and their chain
conformations.

The processing routes to \dph\ and \bph\ are well established.
Poling $\alpha$ at fields $E$ of $0.10$--$0.25$~GV\,m$^{-1}$ converts
it to $\delta$ \cite{davis1978,naegele1978,bachmann1980,li2013,martin2017}.
$\beta$ is obtained by drawing $\alpha$ uniaxially along the chain axis
(lattice constant $b$ in the present convention), which produces the
TTTT conformation, and then poling to align the chain dipoles
\cite{lovinger1983,kepler1992}.
Conversion of $\alpha$ to $\beta$ by field alone is reported only at
substantially higher fields, of order $0.5$~GV\,m$^{-1}$
\cite{davis1978}.
Davis \emph{et al.} described the poling of $\alpha$ as a two-stage
process \cite{davis1978}.
Near $0.10$~GV\,m$^{-1}$ the chain dipoles align with little if any
change of molecular conformation; only near $0.5$~GV\,m$^{-1}$ does
$\beta$ appear.
The present calculation separates those two stages.

First-principles work has established the $E=0$ energetics of the
polymorphs
\cite{nakhmanson2004,ranjan2007,pelizza2016,pelizza2019,itoh2014,xu2024},
the energy barriers along candidate transformation paths
\cite{ranjan2012,kim2016,geng2025}, and the lowering of those barriers
by an applied field \cite{ranjan2012,geng2025}.
In those studies the field enters through an enthalpy term
$-\bm{E}\cdot\bm{P}$, with $\bm{P}$ the spontaneous polarization.
What is missing is the thermodynamics of the transformations under a
controlled electrical boundary condition, that is, the free energy as
a function of the electric displacement $D$ rather than of $E$,
for reasons set out next.

The distinction matters because $E$ and $D$ are different control
variables \cite{stengel2009,hong2011}.
Their two limits correspond to different physics.
Short circuit ($E=0$) admits any spontaneous polarization, whereas
open circuit ($D=0$) imposes a depolarizing field that usually
suppresses it \cite{garrity2014,qiu2021,qiu2023}.
$D$ is also the quantity fixed in a capacitor by the free charge on
the electrodes.
The free energy as a function of $D$ yields the field $E(D)$ directly.
The relation $E(D)$ yields both the stored energy $\int E\,dD$ and the 
coexistence and hysteresis behavior that determine the performance 
of a dielectric energy-storage device \cite{chu2006,ranjan2012}.
Reference~\cite{qiu2026} showed that the fixed-$D$ free energy can be
obtained by post-processing the internal energy, polarization, volume,
and electronic dielectric constant of each atomic configuration.
Applied to LiNbO$_3$, the approach also showed that phases generated
by different soft modes respond to $D$ in qualitatively different
ways, according to the mode effective charges of those modes.

PVDF poses a different and, we believe, instructive problem for this
framework.
First, its phases are not generated by soft phonons of a common parent
structure.
They are connected instead by large-amplitude orientational and
conformational coordinates, along which the polarization can be a
strongly nonlinear function of the coordinate and the cell can strain
by about ten percent.
The configurations must therefore come from constructed transformation
paths.
Second, without a soft mode there is no mode effective charge to
measure how strongly a coordinate couples to $D$.
Third, the polarization quantum of the modern theory of polarization
\cite{kingsmith1993,resta1994} is comparable to the polarization
itself, so the Berry phase of an isolated configuration cannot be read
at face value.
Its branch must be tracked along a continuous path, and here the
transformation path itself supplies that connection.

Here we construct two constrained transformation coordinates in a
common 24-atom cell: a tilt of the gauche C--C bonds that maps
$\alpha$ onto $\delta$, and a backbone torsion that maps $\delta$ onto
$\beta$ with the chain repeat following its natural value.
Along both we compute, with a van der Waals density functional, the
internal energy, the branch-resolved Berry-phase polarization, the
cell volume, and the clamped-ion dielectric tensor.
From these we obtain the fixed-$D$ free-energy landscapes and the $D$
dependence of both barriers.
To find what sets their different responses, we introduce a path
effective charge, the dipole created per unit atomic displacement,
and test its role by scaling the polarization change along each path.
We then construct the phase sequence and the Maxwell construction for
phase coexistence.
Finally, we test these results against the lattice screening obtained
from the zone-center phonons and Born charges of the relaxed phases.

A finite $D$ reorders the three energy wells much more strongly than it
lowers the saddles between them, because each saddle has a
polarization close to its own critical displacement.
Within that common picture the two steps end differently.
$D$ removes the tilt barrier, which falls below $k_BT$ at room
temperature by $D\simeq0.3$~\Cm.
The torsion barrier survives, so $\delta\to\beta$ stays first order
and hysteretic throughout the range studied.
Much of the difference traces to the path effective charge, which is
about $1.5$ times larger for the tilt.
That split is the thermodynamic content of the empirical route to
$\beta$-PVDF: a field produces $\delta$, and drawing produces $\beta$.
It also suggests a chemical route to field-driven $\beta$: raising the
path effective charge of the conformational step.

\section{Theory and methods}

\subsection{Free energy at fixed \texorpdfstring{$D$}{D}}
We use the free energy of Ref.~\cite{qiu2026}, summarized here for
completeness.
For an atomic configuration $\lambda$, the Legendre transform of the
fixed-$E$ free energy \cite{souza2002,fu2003} with respect to the
conjugate pair $(E,D)$ yields
\begin{equation}
\Phi(\lambda;D)=\UKS(\lambda)+\Udp(\lambda;D),
\label{eq:phi}
\end{equation}
with
\begin{equation}
\Udp(\lambda;D)=\frac{\Omega(\lambda)\,[D-P(\lambda)]^2}{2\varepsilon_0[1+\chi_\infty(\lambda)]},
\label{eq:udp}
\end{equation}
where $\Omega(\lambda)$, $\UKS(\lambda)$, $P(\lambda)$, and
$\chi_\infty(\lambda)=\varepsilon_\infty(\lambda)-1$ are the cell
volume, the internal (Kohn--Sham) energy, the polarization along the
field, and the clamped-ion high-frequency dielectric susceptibility
along the same direction.
$\Phi$ is the zero-temperature electric free energy at fixed $D$, the
counterpart of the electric enthalpy used at fixed $E$
\cite{stengel2009}.
Its second term, $\Udp$, is the depolarization energy.
Expanding $(D-P)^2$ gives a $D^2$ field energy, a $-DP$ interaction,
and a $P^2$ self-energy; at $D=0$, $\Udp$ reduces to the familiar
open-circuit depolarization energy.
Unless stated otherwise, $\varepsilon_\infty$ denotes this clamped-ion
constant along the polar axis and carries no component index; a
superscript denotes the phase, as in Eq.~(\ref{eq:parabola}).
Table~\ref{tab:phases} is the exception: there the superscripts label
Cartesian components.

The macroscopic field in configuration $\lambda$ is
$E(\lambda;D)=[D-P(\lambda)]/\varepsilon_0[1+\chi_\infty(\lambda)]$.
At a configuration that minimizes $\Phi$ over $\lambda$, where
$\partial\Phi/\partial\lambda=0$, the same $E$ is the total derivative
$\Omega^{-1}\,d\Phi_{\min}/dD$.
Equation~(\ref{eq:phi}) is a post-processing expression: $\UKS$,
$\Omega$, $P$, and $\chi_\infty$ are computed once for each
configuration at $E=0$, and $\Phi$ is then evaluated for any $D$.
It complements the direct constrained-$D$ method \cite{stengel2009},
which minimizes forces and stresses at a given $D$, by giving the
response of each phase and of the whole transformation path.
This post-processing rests on two assumptions.
Electronic screening of the depolarizing field is described by
$\varepsilon_\infty$, and the relevant ionic response is contained in
$\lambda$ itself.
Section~\ref{sec:screening} tests the second assumption by evaluating
the lattice response of the modes orthogonal to $\lambda$, and finds
it small.

In Ref.~\cite{qiu2026} the configurations were generated by freezing
in soft-phonon eigendisplacements of a paraelectric reference
structure.
Here they are the members of two constrained transformation paths,
described next, and $\lambda$ stands for the tilt angle $t$ on the
first path and for the gauche torsion angle $\varphi$ on the second.
We keep the two physical names rather than an indexed $\lambda$
because the two coordinates move different things
(Sec.~\ref{sec:coords}).

\subsection{Transformation coordinates}
\label{sec:coords}

\begin{figure*}[tb]
\includegraphics[width=0.8\textwidth]{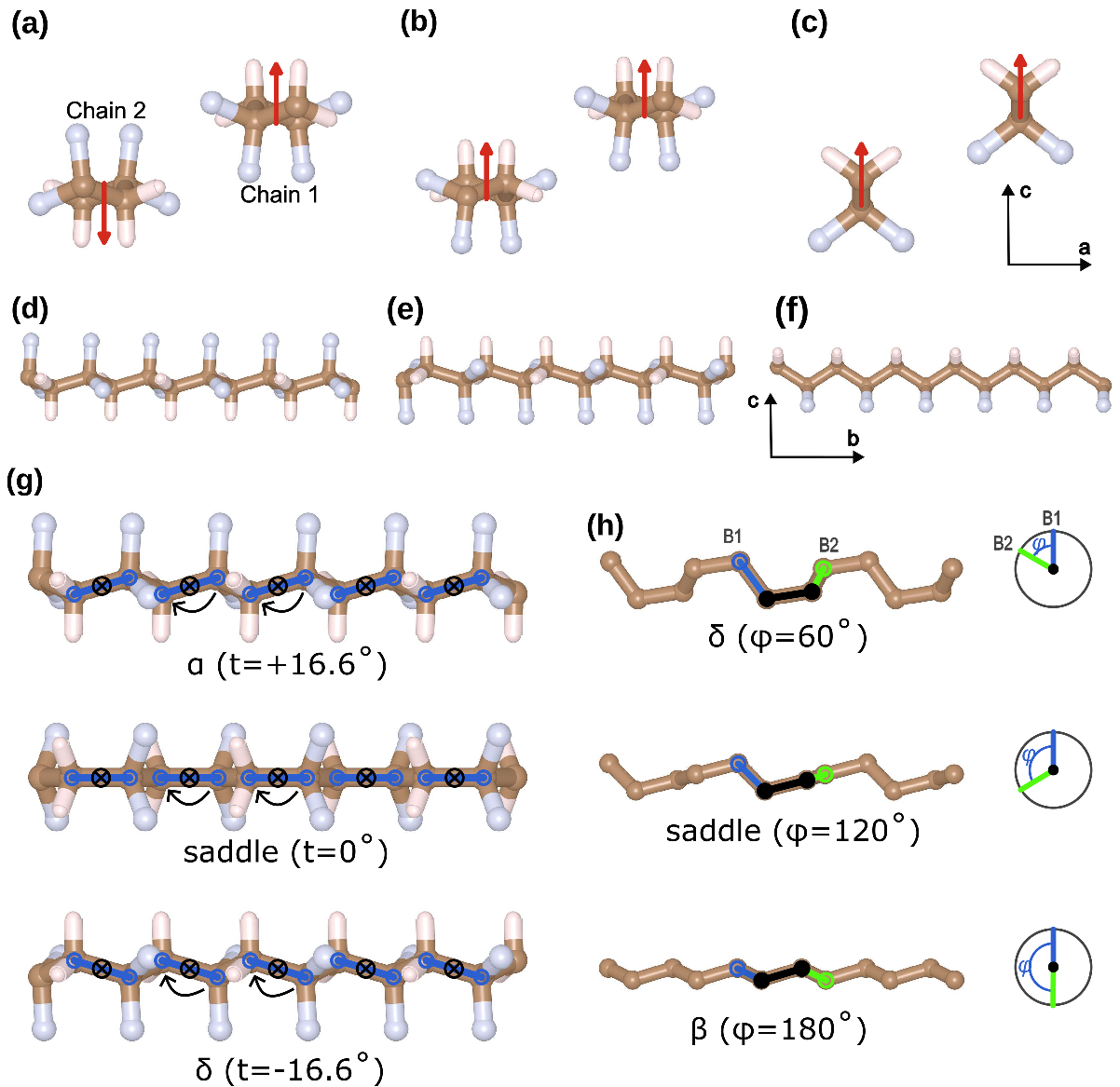}
\caption{The three polymorphs and the two coordinates in the common frame
($b$ = chain axis, $c$ = polar axis; C brown, F pale blue, H pink).
(a)--(c) \aph, \dph, and \bph\ viewed along $b$; red arrows are the chain
dipoles, which cancel in \aph\ and add in \dph\ and \bph.
(d)--(f) Chain 2 of each phase viewed along $a$: \aph\ and \dph\
share the TG$^+$TG$^-$ conformation, and \bph\ is all-trans.
(g) The \aph$\to$\dph\ tilt of chain~2, viewed along $a$: relaxed \aph\
($t=+16.6^\circ$), the saddle ($t=0$), and relaxed \dph\ ($t=-16.6^\circ$).
Each blue gauche C--C bond (two per repeat, carbons ringed) rotates
about an axis parallel to $a$ through its center ($\otimes$, into the
page); curved arrows give the sense, and over the path each bond turns
by $-33.2^\circ$.
(h) The \dph$\to$\bph\ torsion
(backbone only, oblique view about $40^\circ$ from $c$;
rows scaled to equal repeat).
$\varphi$ is the backbone torsion about the
central C--C bond (black), between B1 (blue) and B2 (green): relaxed \dph\
($\varphi=60^\circ$), the saddle ($\varphi=120^\circ$), and relaxed \bph\
($\varphi=180^\circ$).
Circles are Newman projections down that bond, sighted from the B1 end.}
\label{fig:struct}
\end{figure*}

All three phases are described in a single 24-atom orthorhombic cell
containing two chains of two monomers each, with the chain axis along
$b$ and the polar axis along $c$ (Fig.~\ref{fig:struct}).
Two chains are the minimum that distinguishes \aph\ from \dph, which
differ in the relative orientation of neighboring chains.
\aph\ and \dph\ are primitive in this cell, and \bph\ (six atoms per
primitive cell) is commensurate with it.
A common cell and a common polar axis make $D$ a single control
variable for the whole pathway.
Throughout, $P_c$ denotes the polarization along $c$.
By symmetry it is the only nonzero component in the polar phases
\dph\ and \bph, and it is the component to which $D$, also along $c$,
couples.

The $\gamma$ phase (space group $Cc$; T$_3$G$^+$T$_3$G$^-$) is
omitted.
Its primitive cell contains 48 atoms, its polarization lies
$33^\circ$ off $c$ in the present frame, and its C-centered lattice
admits no orthogonal primitive cell.
It therefore cannot share the common cell.
This restricts the subspace; it does not imply that $\gamma$ is
irrelevant or that a path through it is forbidden.
$\gamma$ is reached thermally or from solution \cite{lovinger1983},
and neither route is treated here.

\emph{\aph$\to$\dph: tilt of the gauche bonds.}
The two phases have the same TG$^+$TG$^-$ chain conformation
(backbone torsions $180/{-}60/{-}180/{+}60^\circ$ in both, equal to
within $0.4^\circ$) and differ only in the setting of the second
chain: antiparallel dipoles in $\alpha$, parallel dipoles in $\delta$.
The literature describes the conversion as a rigid $180^\circ$
rotation of one chain about its axis \cite{kim2016,geng2025}.
We use a smaller coordinate.
Of the four C--C bonds per repeat, the two gauche bonds (those about
which the torsion is gauche) lie in the $b$--$c$ plane
[Fig.~\ref{fig:struct}(g)].
We treat each as a rigid unit and define the tilt $t$ as its angle
from $b$ in that plane.
Rotating both units of chain~2 about axes parallel to $a$ through
their own bond centers, from $t=+16.6^\circ$ (\aph) to
$t=-16.6^\circ$ (\dph), together with a rigid shift of chain~2 by
$0.28$~\AA\ (mostly along $c$), maps the carbon skeleton of chain~2 in
\aph\ onto that of \dph\ with a root-mean-square deviation of
$9\times10^{-4}$~\AA.
Along the path the two units of a repeat also move apart along $a$,
by $0.12$~\AA\ at the midpoint, so that the trans C--C bonds keep
their length.
The midpoint $t=0$ is the saddle: there the gauche bonds lie along
$b$ and the dihedral about them passes through $0^\circ$.
Nine intermediate configurations are generated at equal steps in $t$.

The F and H atoms cannot be carried rigidly with either unit.
Each carbon has two backbone neighbors that belong to different rigid
units, so its CH$_2$ or CF$_2$ frame is not fixed by a single unit.
The substituents were therefore rebuilt from the local backbone frame
at each step and then relaxed.
The relaxed $\alpha$ and $\delta$ cells differ by only $-0.81\%$ in
$a$, $+0.06\%$ in $b$, and $+0.08\%$ in $c$ (Table~\ref{tab:phases}).
The cell is interpolated linearly in $t$ between them, so the first
and last images are exactly the relaxed $\alpha$ and $\delta$
structures.
With the atoms held in fractional coordinates, this cell change is an
affine strain that alters the C--C bond lengths by less than
$0.008$~\AA.
Repeating the scan in a fixed $\alpha$ cell lowers the barrier by
$12$~meV per cell ($2\%$), which we take as the cell-path uncertainty
of the tilt barrier.

Applying the tilt to chain~2 rather than chain~1 selects one of the two equivalent
\dph\ domain states: the inversion operation of \aph\ maps one choice onto
the other and reverses the sign of $P$. We orient $+c$ along the polarization
that the path produces, so that $P_\delta$, $P_\beta$ and the values of $D$
considered below are positive. The opposite choice reverses $P$, $D$ and $E$
together and changes nothing else. The two choices are the $\pm P_\delta$
states whose tie line forms the hull at $|D|<P_\delta$
(Sec.~\ref{sec:sequence}).

\emph{\dph$\to$\bph: backbone torsion with $b$ slaved to $\varphi$.}
Here the conformation changes.
The control parameter $\varphi$ is the magnitude of the dihedral about a
gauche bond, the angle between the two trans bonds that flank it
[B1 and B2 in the Newman projections of Fig.~\ref{fig:struct}(h)].
On each chain the two gauche dihedrals open in opposite senses, from
$\pm60^\circ$ ($\delta$) to $\pm180^\circ$ ($\beta$) in ten equal steps, so
the chain acquires no net twist and its glide plane is preserved along the
whole path.
The chain repeat (lattice constant $b$) is slaved to $\varphi$ and
rises from $4.713$ to $5.213$~\AA\ ($+10.6\%$).
Holding $b$ at the $\delta$ value while opening $\varphi$ would close
the C--C--C angle to $99^\circ$ at $\varphi=180^\circ$, against
$115^\circ$ in relaxed $\beta$.
We therefore set $b(\varphi)$ to the natural repeat of the chain at
each step and relax $a$ and $c$ at that fixed $b$.
The torsion and the uniaxial strain along the chain axis are treated
as a single constrained coordinate.

The torsion cannot reverse a chain.
Opening the two gauche dihedrals in opposite senses preserves the
glide plane, so each chain dipole stays in the $b$--$c$ plane.
Its $c$ projection grows monotonically from the TG$^+$TG$^-$ value $p$
to about $2p$ (Sec.~\ref{sec:db0}); on the tilt, by contrast, it
passes through zero at the cis saddle (Sec.~\ref{sec:ad0}).
Applied to $\alpha$, the torsion would therefore give not $\beta$ but
the antipolar all-trans $\zeta$ phase of Ref.~\cite{pelizza2019}, the
``nonpolar $\beta$'' of route~II in Ref.~\cite{kim2016}.
One chain must be reversed, before the torsion or after it.
Before it, a TG$^+$TG$^-$ chain is reversed by the tilt used here or
by a rigid rotation about its axis \cite{ranjan2012,kim2016,geng2025}.
After it, the all-trans chain must be rotated about its axis; that
route is not treated here.

Both paths are constrained scans, not nudged elastic bands. At each
configuration the carbon skeleton is placed by construction and frozen, and
all F and H atoms relax. Each scan begins and ends on an independently
relaxed phase.
The two scans share the relaxed \dph\ exactly. The barriers so obtained are
upper bounds in two senses: the skeleton is confined to a chosen coordinate,
and the 24-atom cell forces every repeat unit to transform in step.
Nucleation-and-growth mechanisms \cite{dveyaharon1980} are excluded by
construction.

\emph{Two coordinates, not one.}
We deliberately do not combine the two scans into a single reaction
coordinate, because they are fundamentally different.
The tilt acts on one chain and leaves its conformation unchanged at
the end points; the cell is invariant to within $1\%$.
The torsion acts on both chains, converts TG$^+$TG$^-$ into TTTT, and
strains $b$ by $10.6\%$.
$D$ also acts on the two coordinates differently
(Sec.~\ref{sec:landscape}).
The scans are used together in only two places: (i) the branch
tracking of the Berry phase, which needs a continuous chain of
configurations from centrosymmetric \aph\ through \dph\ to \bph\
(Sec.~\ref{sec:branch}), and (ii) the assignment of the minima of
$\Phi$ to the \aph, \dph, and \bph\ basins, which are separated by the
$\UKS$ saddles of the two scans (Sec.~\ref{sec:sequence}).
All other results are presented per coordinate.

\subsection{Branch tracking of the Berry-phase polarization}
\label{sec:branch}

The polarization along $c$ is defined modulo the quantum
$P_{\rm q}=e/ab$ \cite{kingsmith1993,resta1994}.
For the present cell it is $0.353$~\Cm\ in \aph\ and varies between
$0.356$ and $0.341$~\Cm\ along the \dph$\to$\bph\ path as $a$ shrinks
and $b$ grows.
Because $|P_\beta|=0.194$~\Cm\ is more than half of $P_{\rm q}$, the
raw Berry phases of the individual configurations fall on several
branches and are meaningless at face value
[Figs.~\ref{fig:ad0}(b) and \ref{fig:db0}(b)].
We fix the branch by continuity.
$\alpha$ is centrosymmetric, so $P_\alpha=0$ by symmetry, and the two
scans form a continuous adiabatic path from $\alpha$ through $\delta$
to $\beta$.
Each configuration is assigned the branch nearest its predecessor,
using the polarization quantum of its own cell.
Unlike the usual procedure, which interpolates from a nonpolar
reference structure \cite{pelizza2016,qiuBTO2023}, this one needs no
artificial reference: the physical path supplies the connection.

Two checks support the assignment.
(i) In $\alpha$ it gives $|P|=3\times10^{-6}$~\Cm, $10^{-5}$ of the
quantum.
(ii) The ratio $|P_\beta|/|P_\delta|=2.01$ matches the projection of
the CF$_2$ dipoles onto $c$ in the two conformations
(Sec.~\ref{sec:db0}).
A separate calculation of $\beta$ in its own relaxed cell shows what
is at stake.
Its raw Berry phase is $-0.162$~\Cm.
With the quantum $0.356$~\Cm\ of that cell the admissible values are
$-0.517$, $-0.162$, and $+0.194$~\Cm, and continuity from $\alpha$
selects the last.
(The scan end point in Fig.~\ref{fig:db0}(b) happens to fall on the
physical branch.)
The branch nearest zero would have given $0.162$~\Cm\ and a ratio of
$1.68$, a $17\%$ error that looks plausible in isolation.

\subsection{Computational details}

Calculations were performed with \textsc{Quantum ESPRESSO}
\cite{giannozzi2009,giannozzi2017,giannozzi2020} using the vdW-DF
exchange-correlation functional \cite{dion2004} in its self-consistent
implementation \cite{thonhauser2007,thonhauser2015}; the method is
reviewed in Refs.~\cite{langreth2009,berland2015}.
The functional describes the dispersive interchain cohesion of the
molecular crystal, and its lattice parameters for the PVDF polymorphs
agree with experiment to about $2\%$ \cite{pelizza2016}.
Pseudopotentials were taken from the SSSP Efficiency~1.3 library
\cite{prandini2018} (PAW for C, ultrasoft for F and H).
Plane-wave cutoffs of 90 and 720~Ry for wave functions and density
and a $2\times4\times4$ Monkhorst--Pack mesh were fixed by a
convergence study on a 12-atom \bph\ cell.

Cells were relaxed with all three lattice parameters free and the
angles held at $90^\circ$, using the Broyden--Fletcher--Goldfarb--Shanno
algorithm and the vdW-DF stress tensor \cite{sabatini2012}.
Relaxations were converged when the total energy changed by less than
$10^{-5}$~Ry, every force component fell below
$10^{-4}$~Ry\,bohr$^{-1}$, and every stress component fell below
$0.1$~kbar.
The self-consistency threshold was $10^{-9}$~Ry, tightened to
$10^{-12}$~Ry for Berry-phase and perturbation-theory calculations.
After relaxation the cell was held fixed and a fresh self-consistent
calculation was performed before any energy was quoted.
Holding the angles at $90^\circ$ constrains \aph, whose monoclinic
angle is free by symmetry and relaxes to $90.6^\circ$ with the same
functional \cite{pelizza2016}.
The energy cost of that shear is negligible on the scale of
Table~\ref{tab:phases}.

In the constrained scans the F and H atoms were relaxed at fixed
carbon skeleton with the same force and energy thresholds.
Polarization was computed by the Berry-phase method with ten $k$
points per string along $c$.
The clamped-ion dielectric tensor, the Born effective charges, and the
zone-center phonons of the three relaxed phases were obtained from
density-functional perturbation theory \cite{baroni2001,gonze1997}.
From these Born charges and force constants we evaluated the lattice
contribution to the dielectric constant and its part orthogonal to
each transformation coordinate (Sec.~\ref{sec:screening}).

\section{Results}

The first three subsections are at $E=0$ (short circuit), where $D=P$
and the depolarization term of Eq.~(\ref{eq:phi}) vanishes.
$D$ becomes the control variable from Sec.~\ref{sec:landscape} onward.

\subsection{Polymorphs at \texorpdfstring{$E=0$}{E = 0}}\label{sec:phases}

Table~\ref{tab:phases} collects the relaxed structures.
Every lattice parameter lies within $0.7\%$ of the previous vdW-DF 
result \cite{pelizza2016}.
The residual disagreement with experiment is therefore that of the
functional, not of convergence.
Against the experimental values in Table~\ref{tab:phases} the largest
 deviations are about $2\%$.
The calculated chain axis is longer than experiment in both \aph\ 
and \bph.
That overestimate is shared by the phases and largely cancels in 
phase differences.
In \bph\ the calculated $c$ is also shorter than experiment.
The energetic ordering agrees with previous vdW-DF and PBE results in
placing \dph\ marginally below \aph\ \cite{pelizza2016}, and with the
near degeneracy of the two phases found with other functionals
\cite{kim2016,geng2025}.
Our difference, $-0.34$~kJ\,mol$^{-1}$ ($-3.6$~meV) per monomer,
compares with $-0.3$~kJ\,mol$^{-1}$ from vdW-DF \cite{pelizza2016}.
A difference of that size is comparable to zero-point and vibrational
free-energy differences, so the static totals do not decide the
finite-temperature ordering of the two phases
(Sec.~\ref{sec:discussion}).
What matters below is that the packing energy that separates
antiparallel from parallel TG$^+$TG$^-$ chains is small.

Our $\beta$--$\alpha$ internal-energy difference,
$+1.89$~kJ\,mol$^{-1}$ per monomer, is about half the
$\approx+3.8$~kJ\,mol$^{-1}$ obtained from Table~4 of
Ref.~\cite{pelizza2016} with the same functional.
The factor of two is less alarming than it seems.
The $\beta$--$\alpha$ difference is a small residual of two large
terms of opposite sign: the intrachain cost of straightening
TG$^+$TG$^-$ into TTTT ($+17.6$~kJ\,mol$^{-1}$ per monomer in
Ref.~\cite{pelizza2016}) and the stronger interchain cohesion of the
denser $\beta$ packing ($-13.8$~kJ\,mol$^{-1}$).
A $10\%$ change in either term, of the size produced by different
pseudopotentials or cutoffs, shifts the residual by
$1.4$--$1.8$~kJ\,mol$^{-1}$, as much as the discrepancy itself.

Experiment supplies a bound that discriminates between the two residuals.
The field at which $\alpha$ and $\beta$ have equal electric enthalpy is
$E^*_{\alpha\beta}=\Delta U_{\beta\alpha}/(\Omega P_\beta)$.
That field is a thermodynamic lower bound on the field at which
$\beta$ appears from $\alpha$ films,
$\simeq 0.5$~GV\,m$^{-1}$ \cite{davis1978}.
Rearranging gives
$\Delta U_{\beta\alpha}\le E_{\rm meas}\,\Omega P_\beta
\simeq 3.2$~kJ\,mol$^{-1}$ per monomer.
The present value lies below this limit; the published residual
lies slightly above it.
Adopting the published residual would not change the fixed-$D$
sequence and would move $D_{c2}$ only from $0.157$ to about
$0.17$~\Cm, but it would nearly double $E^*_{\delta\beta}$, to about
$1.3$~GV\,m$^{-1}$, because $E^*$ is proportional to $\Delta U$
(Sec.~\ref{sec:sequence}).

All three phases are dynamically stable at the zone center: none of
the 72 modes of each cell is imaginary, and the three acoustic modes
vanish after the sum rule.
In particular, $\delta$ is a local minimum of the functional, not a
saddle.
The lowest optic mode of $\beta$ ($13.7$~cm$^{-1}$) is markedly softer
than those of $\alpha$ and $\delta$.
The clamped-ion dielectric constants are $2.1$--$2.4$, largest along
the covalent chain direction, and compare well with the optical
dielectric constant of $\beta$-phase P(VDF--TrFE) films,
$n^2\simeq 2.0$ ($n\simeq 1.42$ \cite{wang2024}), the nearest measured
counterpart.

The branch-resolved spontaneous polarization of \bph, $0.194$~\Cm,
agrees to $2\%$ with the vdW-DF value of Ref.~\cite{pelizza2016}
($0.198$~\Cm) and lies within the range of other functionals
($0.17$--$0.27$~\Cm\ \cite{pelizza2016,itoh2014,nakhmanson2004}).
Measured polarizations of poled films and oligomer crystals
($0.076$--$0.13$~\Cm) are lower, as expected for finite crystallinity
and imperfect alignment \cite{kepler1992,furukawa1989}.
The polarization of \dph, $0.096$~\Cm, is half that of \bph\
(Sec.~\ref{sec:db0}).
Electroformed \dph\ films give a remanent polarization of about
$0.07$~\Cm\ \cite{li2013}, a larger fraction of the crystal value than
for \bph.
Those films have a strong in-plane chain texture, so most chains lie
perpendicular to the field \cite{li2013}.
Even so, the crystallinity of about $50\%$ assumed in
Ref.~\cite{li2013} would limit the crystalline contribution to
$0.048$~\Cm.
Either the films are more crystalline than assumed, or the functional
underestimates $P_\delta$; other functionals give
$0.086$--$0.14$~\Cm\ \cite{pelizza2016}.

\subsection{\texorpdfstring{$\alpha\to\delta$ at $E=0$}{alpha to delta at E = 0}: an electrical reorientation}
\label{sec:ad0}

\begin{figure*}[tb]
\includegraphics[width=4in]{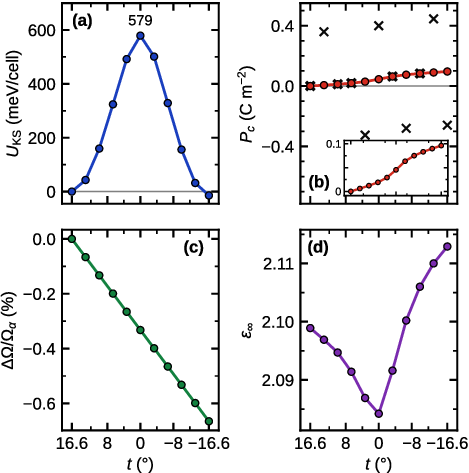}
\caption{The \aph$\to$\dph\ tilt at $E=0$, from relaxed \aph\
($t=+16.6^\circ$) to relaxed \dph\ ($t=-16.6^\circ$).
(a) Internal energy relative to \aph; the fixed-$D$ barrier at $D=0$
is $0.66$~eV (Table~\ref{tab:barriers}).
(b) Polarization along $c$: raw Berry phases (crosses) and the
\aph-continuous branch (circles); the inset enlarges the branch.
(c) Volume change; the cell is interpolated linearly between the
relaxed end points.
(d) Clamped-ion dielectric constant along $c$.}
\label{fig:ad0}
\end{figure*}

Figure~\ref{fig:ad0} shows the $E=0$ properties along the tilt.
The step is an almost purely electrical reorientation: the cell
changes by less than $1\%$ in any dimension [Table~\ref{tab:phases};
volume in Fig.~\ref{fig:ad0}(c)].
What turns is the dipole of a whole chain.
In $\alpha$ the dipoles of the two TG$^+$TG$^-$ chains are equal and
opposite, so the cell is nonpolar by cancellation.
The tilt moves only chain~2 and leaves both chains in the
TG$^+$TG$^-$ conformation at the two ends.
The CF$_2$ group is also rigid: its C--F bonds and F--C--F angle vary
by less than $1\%$ among the three phases and both saddles.
The chain dipoles of $\delta$ therefore have the same magnitudes as
those of $\alpha$ and differ only in direction.
The dipole of chain~2 rotates in the $b$--$c$ plane until its $c$
component has reversed sign, and its $b$ component ends at the
magnitude it had in $\alpha$.
In $\delta$ the two $c$ components add to give $P_\delta$, while the
$b$ components still cancel.

The barrier is $579$~meV per cell ($72.4$~meV per carbon atom),
reached at the eclipsed configuration $t=0$ (TCTC, with C the
eclipsed cis form; Sec.~\ref{sec:coords}).
There chain~2 is eclipsed while chain~1 keeps the $\alpha$ tilt of
$+16.6^\circ$, so the saddle is not a symmetric pair of opposite tilts
and is already polar.

The polarization follows an S-shaped curve from zero in $\alpha$ to
$0.096$~\Cm\ in $\delta$ [Fig.~\ref{fig:ad0}(b) and inset].
It changes slowly near the ends and fastest at the saddle, where it is
already $0.045$~\Cm, $47\%$ of $P_\delta$.
That is the kinematics of a rotating dipole: $P_c\propto\mu\cos\theta$,
with $\mu$ the dipole of chain~2 and $\theta$ its angle from $c$,
changes fastest when the dipole lies along $b$ ($\theta=90^\circ$).
The crosses in Fig.~\ref{fig:ad0}(b) are raw Berry phases that jump
between branches; only the branch continuous from $\alpha$ (circles)
is physical (Sec.~\ref{sec:branch}).
The clamped-ion constant $\varepsilon_\infty$ has a shallow minimum at
the saddle, $2.084$ against $2.099$ in $\alpha$ and $2.113$ in
$\delta$ [Fig.~\ref{fig:ad0}(d)].
It measures the electronic polarizability along $c$ at fixed ions,
which falls when the CF$_2$ groups of chain~2 point along $b$.

The barrier per carbon atom is comparable to values reported for the
same conversion along a different coordinate, a rigid rotation of one
chain about its axis: $>73$~meV per carbon for pure PVDF in
Ref.~\cite{geng2025} (less for the copolymers) and $50$~meV per carbon
in Ref.~\cite{ranjan2012}.
All three calculations are concerted: every repeat unit of the
transformed chain moves at once, so the barrier scales with the length
of chain that is converted rather than with a local activated region.
Here and in Ref.~\cite{ranjan2012} the cell enforces this: with two
monomers per chain, the repeat units are periodic images and cannot
move independently.
In Ref.~\cite{geng2025} the chain is longer, twelve backbone carbon
atoms, but the path is prescribed by a single interchain angle and a
single average gauche dihedral, which imposes the same collective
motion.

Kim \emph{et al.} \cite{kim2016} obtained a much lower barrier:
$21$~meV per monomer ($10$~meV per carbon) for $\alpha\to\delta$, and
$53$~meV per monomer for the highest saddle of their route~I.
Their chain is six monomers long, as in Ref.~\cite{geng2025}, but the
path is a generalized solid-state nudged elastic band that chooses its
own saddle.
The chain can therefore transform one segment at a time, and the cell
can relax along the path.
The factor of $3$--$7$ between that barrier and the concerted values
measures what a nonconcerted mechanism gains.
The barriers of the present 24-atom cell are therefore upper bounds (Sec.~\ref{sec:coords}).

\subsection{\texorpdfstring{$\delta\to\beta$ at $E=0$}{delta to beta at E = 0}: a strain-coupled conformational change}
\label{sec:db0}

\begin{figure*}[tb]
\includegraphics[width=4in]{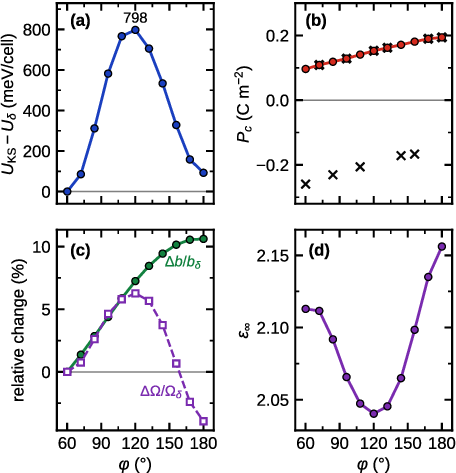}
\caption{The \dph$\to$\bph\ torsion at $E=0$, from relaxed \dph\
($\varphi=60^\circ$) to relaxed \bph\ ($\varphi=180^\circ$).
(a) Internal energy relative to \dph; the fixed-$D$ barrier at $D=0$
is $1.43$~eV (Table~\ref{tab:barriers}).
(b) Polarization along $c$: raw Berry phases (crosses) and the
\aph-continuous branch (circles).
(c) Changes of the chain repeat $b$, which follows the natural repeat
of the chain, and of the volume $\Omega$, relative to \dph; $a$ and
$c$ are relaxed at each $\varphi$.
(d) Clamped-ion dielectric constant along $c$.}
\label{fig:db0}
\end{figure*}

The torsion step (Fig.~\ref{fig:db0}) is of a different kind, because the
chain repeat is part of the coordinate: $b$ grows by 10.6\% as $\varphi$
opens [Fig.~\ref{fig:db0}(c)], while $a$ shrinks by 9.5\% and $c$ by
4.0\% (Table~\ref{tab:phases}).
The barrier of 798~meV per cell (99.7~meV per carbon atom) is reached at
$\varphi=120^\circ$, where the chain is eclipsed.
It lies in the range 61--101~meV per carbon reported for the VDF
copolymers of Ref.~\cite{geng2025} and close to the 100~meV per carbon of
Ref.~\cite{ranjan2012}.

Here the chain dipoles themselves change.
The TG$^+$TG$^-$ kink holds the two CF$_2$ dipoles of a repeat about
$104^\circ$ apart, so their resultant is only $62\%$ of the aligned
value and is tilted off $c$ toward the chain axis $b$.
The cell polarization of $\delta$ is still along $c$, because the $b$
components of the two chains cancel by symmetry.
As $\varphi$ opens to $180^\circ$ the $104^\circ$ angle closes to zero,
the $b$ component of each chain vanishes, and the whole of a larger
chain dipole lies along $c$.
Both chains do this at once, and the cell polarization rises from
$P_\delta$ to $P_\beta$, a factor of $2.01$, nearly linearly in
$\varphi$ [Fig.~\ref{fig:db0}(b)].
Five of the eleven raw Berry phases in that panel lie one quantum
below the physical branch; restored to the $\alpha$-continuous branch
(Sec.~\ref{sec:branch}), they give the monotonic rise shown.

The two coordinates thus recover dipole in different ways.
The tilt removes the cancellation between chains: the $c$
components become parallel while the chain-dipole magnitude is
unchanged.
The torsion completes the alignment within a chain: it closes
the $104^\circ$ angle and folds the leftover $b$ components onto $c$.

The internal energies of the $\delta$ and $\beta$ end points lie much
closer to each other than to the barrier between them.
$\beta$ is only $93$~meV per cell above $\delta$, whereas the eclipsed
maximum is $798$~meV above $\delta$ and $705$~meV above $\beta$.
That high barrier from either side is the $E=0$ origin of the
retention of drawn $\beta$ film: once the all-trans packing has
formed, return to $\delta$ must cross the same eclipsed maximum.
The $93$~meV gap is about half the $170$~meV per cell
($4.1$~kJ\,mol$^{-1}$ per monomer) implied by the chain and cohesive
energies of Ref.~\cite{pelizza2016}.
That is the same factor as for the $\beta$--$\alpha$ difference
(Sec.~\ref{sec:phases}); it reflects how the functional balances the
intrachain and interchain terms, not the path.

The volume maximum is a transient dilation, not a monotonic swelling
[Fig.~\ref{fig:db0}(c)].
The chain repeat $b$ grows monotonically by $10.6\%$, while the
volume rises $6.3\%$ above $\delta$ at the eclipsed saddle and then
falls to $3.9\%$ below $\delta$ in $\beta$.
The extra volume is not supplied along the chain.
$a$ contracts monotonically, whereas $c$ expands by $3.1\%$ to a
maximum at $\varphi=108^\circ$ before contracting to the $\beta$
value.
That transient expansion of $c$, together with the lagging
contraction of $a$, produces the volume peak.
The cell becomes denser than $\delta$ only over the last quarter of
the path, near $\varphi=160^\circ$.
The free volume needed to pass the eclipsed conformation is therefore 
opened perpendicular to the chain at the saddle and recovered on the 
way to $\beta$.

The clamped-ion dielectric constant tracks the density
[Fig.~\ref{fig:db0}(d)].
It falls from $2.113$ in $\delta$ to $2.040$ at $\varphi=120^\circ$
and then rises to $2.156$ in $\beta$, the largest of the three
phases.
That correlation is not accidental.
The cell polarizability along $c$,
$(\varepsilon_\infty-1)\Omega$, varies by only $1.4\%$ over the
path (peak to peak), while $\varepsilon_\infty$ itself varies by
$5.7\%$.
The torsion changes dihedrals and volume, not the chemical units, so
the electronic polarizability of the CF$_2$ and CH$_2$ groups is
essentially unchanged.
$\varepsilon_\infty$ then records the volume: when $\Omega$ rises,
$\varepsilon_\infty$ falls, and vice versa.
The tilt behaves differently.
There the volume change is small and monotonic, yet
$\varepsilon_\infty$ has a minimum at the saddle and
$(\varepsilon_\infty-1)\Omega$ falls by $2\%$.
That is a real loss of polarizability as the dipole of chain~2 turns
away from $c$ (Sec.~\ref{sec:ad0}), not a density effect.

\subsection{Free-energy landscapes at fixed \texorpdfstring{$D$}{D}}
\label{sec:landscape}

\begin{figure*}[tb]
\includegraphics[width=4.5in]{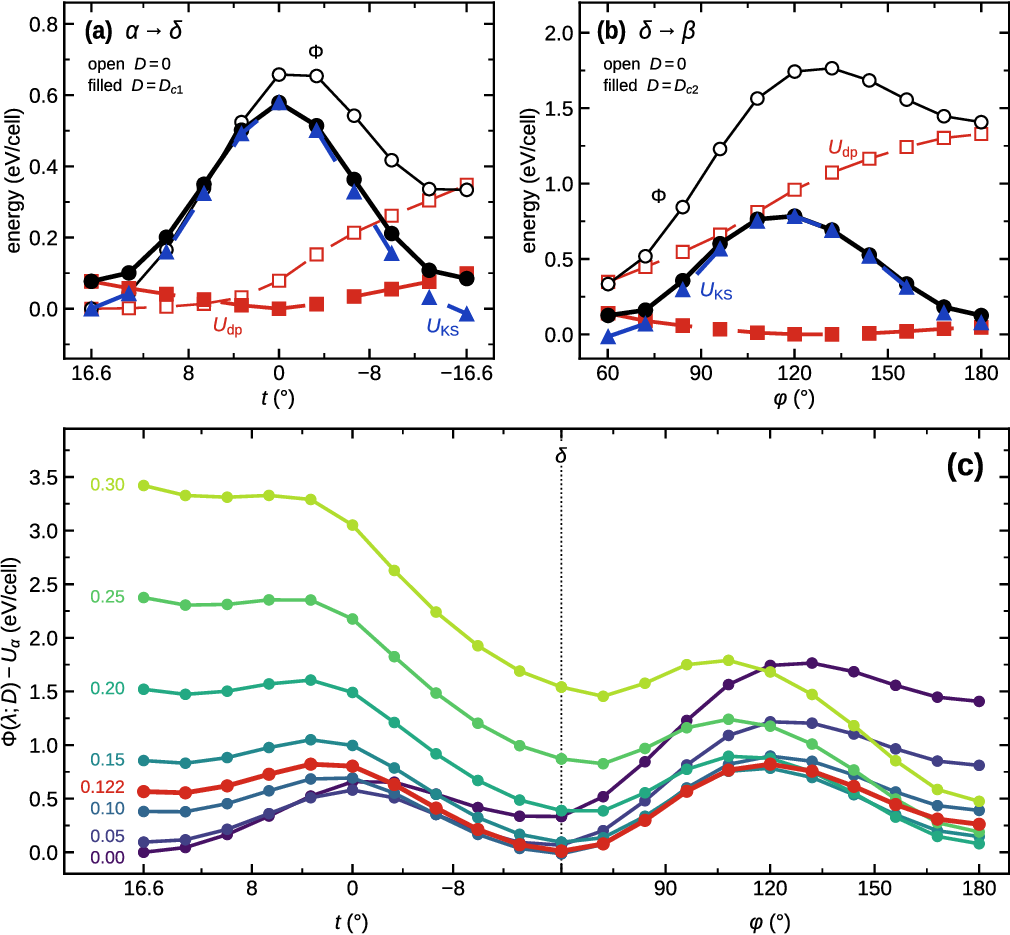}
\caption{Fixed-$D$ free energy along the two coordinates, relative to
$\UKS$ of \aph; $D$ and $P$ point along $c$ in the sense of $P_\beta$.
(a) Tilt and (b) torsion: $\Phi$ (black circles), $\UKS$ (blue
triangles), and $\Udp$ (red squares) at $D=0$ (open) and at the
critical displacement of each step (filled), $D_{c1}=0.046$ and
$D_{c2}=0.157$~\Cm; $\UKS$ does not depend on $D$.
Note the different vertical scales.
(c) $\Phi$ along both coordinates joined at \dph\ (vertical dotted
line), one curve per $D$ (labels in \Cm).
The red curve, $D=0.122$~\Cm, is where the two barriers out of \dph\
are equal, each $0.81$~eV.
All curves are homogeneous states; the multidomain states of
Fig.~\ref{fig:phases}(a) lie below them.}
\label{fig:landscape}
\end{figure*}

Figure~\ref{fig:landscape} shows the free energy of
Eq.~(\ref{eq:phi}) along each coordinate.
Two features of these landscapes control what follows.
First, the depolarization energy is large on the scale of the internal
energy.
At $D=0$, $U_{\rm dp}$ of $\beta$ is $1.33$~eV per cell, which is
just $\Omega_\beta P_\beta^2/2\varepsilon_0\varepsilon_\infty^\beta$
from Table~\ref{tab:phases}.
That is seventeen times the $\beta$--$\alpha$ internal-energy
difference of $78$~meV and $1.7$ times the $E=0$ torsion barrier of
$0.80$~eV.
The corresponding $\Phi$ at $D=0$ is $1.41$~eV.
The cause is the small dielectric constant of the molecular
crystal: $\varepsilon_\infty\simeq 2.2$, or $\simeq 2.6$ with the
lattice response orthogonal to the coordinate
(Sec.~\ref{sec:screening}).

Second, the saddles are dominated by $\UKS$ despite that penalty,
which applies to polar $\beta$ at $D=0$, where $D-P=-P_\beta$.
Each saddle happens to sit almost exactly at its own critical
displacement: the tilt saddle has $P=0.045$~\Cm\ against
$D_{c1}=0.046$~\Cm, and the torsion saddle $P=0.153$~\Cm\ against
$D_{c2}=0.157$~\Cm\ (Sec.~\ref{sec:sequence}).
At those displacement fields $\Udp$ nearly vanishes at the saddle,
and $\UKS$ supplies essentially the full barrier
[Figs.~\ref{fig:landscape}(a) and (b)].
Away from the saddle, $\Udp$ rises smoothly and nearly parabolically
in the coordinate.

A finite $D$ therefore reorders the wells much more than it flattens
the walls between them.
From $D=0$ to $D=0.30$~\Cm, $\Phi_\beta-\Phi_\alpha$ swings by
$4.35$~eV, while the tilt barrier falls by only $0.64$~eV and the
torsion barrier by $1.10$~eV.
The tilt saddle disappears and the torsion saddle does not
[Fig.~\ref{fig:landscape}(c)]: by $D=0.30$~\Cm\ the tilt has lost
$97\%$ of its barrier and the torsion only $76\%$.
Section~\ref{sec:zstar} traces the difference mainly to the larger
path effective charge of the tilt.

The way $D$ acts on each phase follows from Eq.~(\ref{eq:udp}) alone.
The depolarization term vanishes at $D=P_X$ and grows quadratically
away from that point.
At $D=0$ the polar phases therefore sit in their own depolarizing
fields $E=-P/(\varepsilon_0\varepsilon_\infty)$: $-5.1$~GV\,m$^{-1}$
in $\delta$ and $-10.2$~GV\,m$^{-1}$ in $\beta$.
As $D$ increases from zero the initial slopes are
$\mathrm{d}\Phi_\delta/\mathrm{d}D=-7.2$~eV per \Cm\ per cell and
$\mathrm{d}\Phi_\beta/\mathrm{d}D=-13.7$~eV per \Cm\ per cell.
$\Phi_\alpha$, already at $P=0$, can only rise.
Raising $D$ toward $P_X$ removes that penalty; it does not push
$\Phi_X$ below $U_X$.

The depth of each basin below the lowest saddle that bounds it is a separate measure of stability [Table~\ref{tab:barriers}; Fig.~\ref{fig:landscape}(c)].
As $D$ rises from 0 to $0.30$~\Cm, the tilt barrier out of $\alpha$ falls from $0.66$ to $0.02$~eV, so $\alpha$ is steadily destabilized.
Over the same range the reverse torsion barrier out of $\beta$ rises from $0.36$ to $1.32$~eV, so $\beta$ is steadily stabilized and persists once formed.
$\delta$ is bounded on both sides and is neither.
The barrier back to $\alpha$ grows with $D$, while the barrier on to $\beta$ falls.
The two barriers are equal at $D=0.122$~\Cm, where each is $0.81$~eV.
The $\delta$ well never closes.
Across the coexistence window of Sec.~\ref{sec:sequence} the lower of the two barriers stays between $0.50$ and $0.81$~eV, and nowhere in $0\le D\le 0.30$~\Cm\ does it fall below $0.32$~eV, twelve times $k_BT$ at room temperature.
No value of $D$ in this range leaves homogeneous $\delta$ free to leave in either direction.
That is the microscopic content of the hysteresis in Sec.~\ref{sec:barriers}.

The present coordinates support five homogeneous states at $D=0$:
$\alpha$, and the two polar settings of each of $\delta$ and $\beta$
[Fig.~\ref{fig:landscape}(c), $D=0$ curve], the $-P$ partners being
equivalent.
Of these five, $\alpha$ is the lowest; $\pm\delta$ and $\pm\beta$ are
metastable.
Among homogeneous states, therefore, PVDF is not hyperferroelectric
in the sense of Ref.~\cite{garrity2014}: although a metastable polarization
survives at open circuit, the homogeneous ground state is
antipolar.
Allowing the crystal to break into domains changes that conclusion
(Sec.~\ref{sec:sequence}).

\subsection{Barriers and the order of the transitions}
\label{sec:barriers}

\begin{figure*}[tb]
\includegraphics[width=4in]{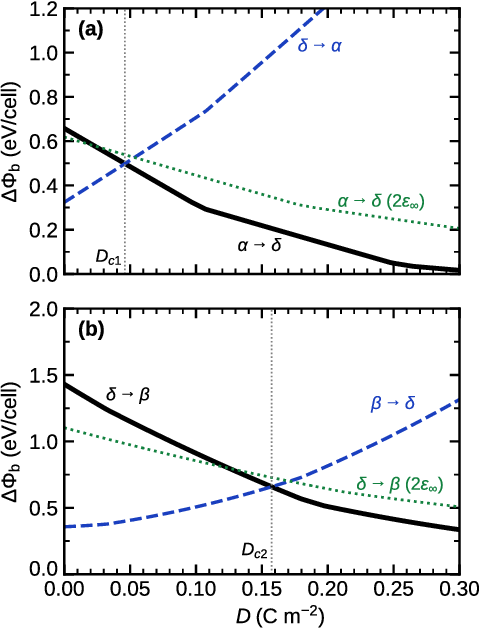}
\caption{Fixed-$D$ barriers along (a) the tilt and (b) the torsion.
$\Delta\Phi_{\rm b}$ is the highest $\Phi$ between two basin minima
along the prescribed coordinate, relative to the starting minimum; it
is not a relaxed saddle.
Solid black, forward; dashed blue, reverse; dotted green, forward with
$\varepsilon_\infty$ doubled (Sec.~\ref{sec:screening}).
Vertical dotted lines mark $D_{c1}$ and $D_{c2}$.
Note the different vertical scales.}
\label{fig:barriers}
\end{figure*}

Figure~\ref{fig:barriers} and Table~\ref{tab:barriers} give the $D$
dependence of the barriers. The two steps behave differently.

The tilt barrier is removed by $D$.
It is $0.66$~eV per cell at $D=0$, larger than the $E=0$ internal-energy
barrier of $0.58$~eV,
because the saddle, whose polarization is $0.045$~\Cm, pays
a depolarization energy that $\alpha$ does not.
The barrier falls to $0.31$~eV at $D=0.10$~\Cm\ and to $0.017$~eV at
$D=0.30$~\Cm.
At that point the $\alpha$ basin has essentially lost its minimum.
The barrier drops below $k_BT$ at room temperature at $D=0.28$~\Cm\
with clamped-ion screening, and at $D=0.35$~\Cm\ with the lattice
screening of Sec.~\ref{sec:screening}.

In this coherent model $D\simeq0.3$~\Cm\ is therefore the limit of
metastability of $\alpha$.
Through $E=(D-P)/\varepsilon_0\varepsilon_\infty$ it corresponds to a
field of order $10$~GV\,m$^{-1}$, an intrinsic field of the model
rather than a predicted coercive field (Sec.~\ref{sec:screening}).
Experimental poling fields are $0.10$--$0.25$~GV\,m$^{-1}$
\cite{davis1978,li2013,martin2017}, two orders of magnitude lower, as
is usual once nucleation and kink propagation \cite{dveyaharon1980}
are allowed; the 24-atom cell cannot describe those.
What the calculation does show is that $D$ couples to the tilt
strongly enough to remove the barrier, so the conversion is not
limited by an intrinsic property of this coordinate.

The torsion barrier is not removed.
It falls from $1.43$~eV per cell at $D=0$ to $0.66$~eV at $D_{c2}$,
where \dph\ and \bph\ are degenerate, and is still $0.34$~eV at
$D=0.30$~\Cm, where the tilt barrier has all but vanished.
Over the whole range the transition is strongly first order: \dph\
keeps a deep metastable minimum far above $D_{c2}$, and \bph\ keeps
one far below it, down to $D=0$, where the reverse barrier is
$0.36$~eV.

The reason is visible in Fig.~\ref{fig:landscape}(b).
The depolarization energy is quadratic in $P-D$ and therefore
smallest where $P\simeq D$.
Near $D_{c2}$ that point is the torsion saddle itself.
$D$ then lowers the barrier only by the depolarization energy it adds
to the $\delta$ end point, $\simeq0.14$~eV, and leaves the $0.8$~eV
conformational and strain cost of the saddle untouched.

\subsection{Why \texorpdfstring{$D$}{D} removes one barrier and not the other}
\label{sec:zstar}

The coupling of a coordinate to $D$ is measured by the dipole it
creates per unit atomic displacement.
For a soft phonon this is the mode effective charge \cite{qiu2026}.
For a finite path we define the analogous quantity
\begin{equation}
Z^*_{\rm path}=\frac{\Omega\,\Delta P}{|\Delta\bm{u}|},
\qquad
|\Delta\bm{u}|=\Big[\sum_{\kappa=1}^{N}\sum_{j=1}^{3}
\big(\Delta u_{\kappa j}\big)^{2}\Big]^{1/2},
\label{eq:zpath}
\end{equation}
where $\Omega\Delta P$ is the change in cell dipole between the end
points and $\Delta\bm{u}$ is the $3N$-component atomic displacement
($N=24$).
Differences are taken in crystal coordinates with the minimum-image
convention and converted to Cartesian with the mean of the two
end-point cells.
Homogeneous strain is thereby excluded, which matters for the torsion,
where $b$ changes by $10.6\%$.
No mass weighting is applied.
The tilt turns one chain ($|\Delta\bm{u}|=4.83$~\AA) and the torsion
unwinds both ($|\Delta\bm{u}|=7.53$~\AA), while both change $P$ by
about $0.098$~\Cm.
Hence $Z^*_{\rm tilt}=0.28e$ and $Z^*_{\rm tors}=0.18e$.

The denominator of Eq.~(\ref{eq:zpath}) is not unique for a
large-amplitude path.
Counting only C and F gives $0.33e$ and $0.23e$, and weighting each
component by $\sqrt{m_\kappa/m_{\rm C}}$ gives $0.27e$ and $0.19e$.
In every convention that keeps F, which carries the dipole,
$Z^*_{\rm tilt}$ exceeds $Z^*_{\rm tors}$ by a factor of
$1.4$--$1.6$.
Because a path is not an eigenvector, $Z^*_{\rm path}$ is not a mode
effective charge, and only the ratio between the two paths is used
below.

\begin{figure*}[tb]
\includegraphics[width=5.5in]{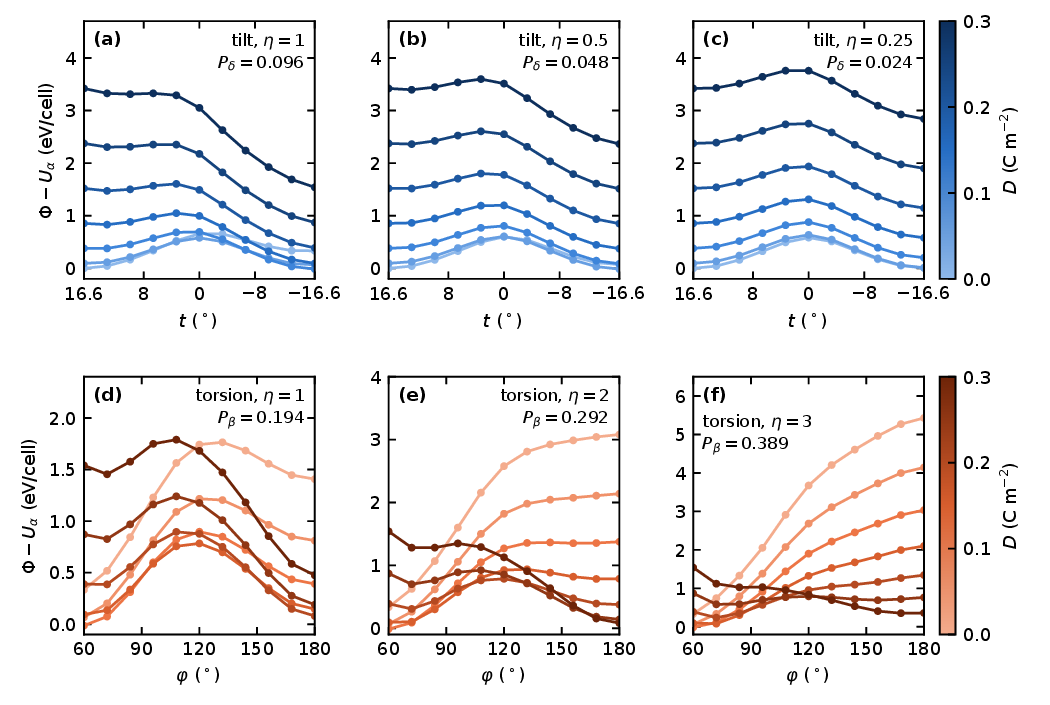}
\caption{Fixed-$D$ free energy with the polarization change along each
path multiplied by $\eta$ at fixed $\UKS$, $\Omega$, and
$\varepsilon_\infty$, which multiplies $Z^*_{\rm path}$ by $\eta$
(Sec.~\ref{sec:zstar}).
(a)--(c) Tilt with $\eta=1$, $0.5$, and $0.25$; (d)--(f) torsion with
$\eta=1$, $2$, and $3$; $\eta=1$ is the calculated path.
Curves run from $D=0$ (light) to $0.30$~\Cm\ (dark) in steps of
$0.05$~\Cm, relative to $\UKS$ of \aph.
Note the different vertical scales.}
\label{fig:zscan}
\end{figure*}

The effective charge sets how strongly $D$ acts on a barrier.
Consider a well of polarization $P_{\rm w}$ and a saddle of
polarization $P_{\rm s}$ with the same
$K=\Omega/2\varepsilon_0\varepsilon_\infty$.
By Eq.~(\ref{eq:udp}), each unit increase of $D$ lowers the barrier
between them by $2K(P_{\rm s}-P_{\rm w})$.
At a fixed atomic displacement, $P_{\rm s}-P_{\rm w}$ grows in
proportion to $Z^*_{\rm path}$.
To isolate this dependence, we multiply the polarization change along
each path by a factor $\eta$ and keep $\UKS$, $\Omega$, and
$\varepsilon_\infty$ fixed.
Along the tilt $P(t)\to\eta P(t)$, and along the torsion
$P(\varphi)\to P_\delta+\eta[P(\varphi)-P_\delta]$.
This multiplies $Z^*_{\rm path}$ by $\eta$; $\eta=1$ is the calculated
path.
Figure~\ref{fig:zscan} shows the resulting free energy for
$0\le D\le0.30$~\Cm.

Reducing the effective charge of the tilt removes the effect of $D$ on
its barrier.
For the calculated path the barrier falls from $0.66$~eV at $D=0$ to
$0.017$~eV at $D=0.30$~\Cm\ [Fig.~\ref{fig:zscan}(a)].
With $\eta=0.5$ it is still $0.20$~eV at $D=0.30$~\Cm, and with
$\eta=0.25$ it is $0.34$~eV [Figs.~\ref{fig:zscan}(b) and (c)].
$D$ then raises the whole profile almost rigidly.
It still lowers $\delta$ relative to $\alpha$, but it no longer
flattens the barrier between them, and $\alpha$ is not converted to
$\delta$ within the range.

Increasing the effective charge of the torsion has the opposite
effect.
With $\eta=2$ the barrier at $D=0.30$~\Cm\ falls to $0.07$~eV, and
with $\eta=3$ to $0.006$~eV, well below $k_BT$ at $300$~K
[Figs.~\ref{fig:zscan}(e) and (f)].
For the calculated torsion it is still $0.34$~eV
[Fig.~\ref{fig:zscan}(d)].
A torsion with a larger effective charge would therefore be flattened
by $D$ in the same way as the tilt.

The larger effective charge has a cost.
It raises $P_\beta$, and the depolarization energy of $\beta$ at
$D=0$ grows as $P_\beta^2$.
For $\eta\ge1.75$, $\beta$ is no longer a local minimum at open
circuit.
With $\eta=2$ it becomes one only above $D=0.09$~\Cm, and with
$\eta=3$ only above $D=0.23$~\Cm\
[the light curves in Figs.~\ref{fig:zscan}(e) and (f)].
The calculated torsion keeps $\beta$ metastable at $D=0$, behind a
reverse barrier of $0.36$~eV (Table~\ref{tab:barriers}).
In this model, the small effective charge that preserves the torsion
barrier under $D$ is thus the same property that lets $\beta$ survive
at open circuit.

Exchanging the two effective charges shows how much of the contrast
they carry.
With the effective charge of the torsion ($\eta=0.64$), the tilt
barrier is still $0.13$~eV at $D=0.30$~\Cm, five times $k_BT$.
The effective charge therefore decides the outcome for the tilt.
With the effective charge of the tilt ($\eta=1.56$), the torsion
barrier at the same $D$ falls from $0.34$ to $0.16$~eV but does not
vanish.
For the torsion the effective charge accounts for about half of the
difference from the tilt.
The rest comes from its higher starting barrier and from its dilated
saddle.
That saddle has a $6.3\%$ larger volume and a smaller
$\varepsilon_\infty$ ($2.04$ against $2.11$) than $\delta$, so its $K$
is $10\%$ larger, whereas $K$ changes by less than $0.5\%$ along the
tilt.

\subsection{Phase sequence and coexistence as a function of \texorpdfstring{$D$}{D}}
\label{sec:sequence}

\begin{figure*}[tb]
\includegraphics[width=3in]{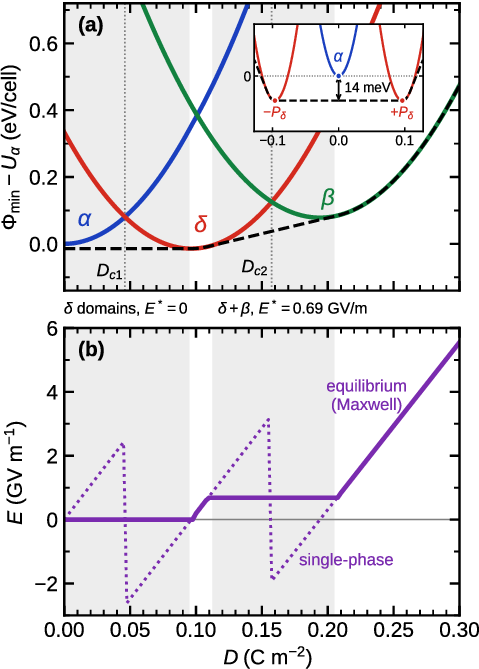}
\caption{(a) Free energy of the local minimum of each basin versus $D$
(solid), relative to $\UKS$ of \aph, and the convex hull over $\pm D$
(dashed).
Vertical dotted lines mark $D_{c1}$ and $D_{c2}$.
Shaded regions have no homogeneous state on the hull: a \dph\
multidomain state for $|D|<P_\delta=0.096$~\Cm\ ($E^*=0$), and
\dph+\bph\ coexistence for $0.11<D<0.21$~\Cm\
($E^*=0.69$~GV\,m$^{-1}$).
Inset: the region near $D=0$, including the $-P_\delta$ branch; the
\dph\ tie line lies $14.3$~meV per cell below homogeneous \aph.
(b) Macroscopic field.
Dotted, $E=(D-P)/\varepsilon_0\varepsilon_\infty$ of the homogeneous
global minimum, which jumps at each crossing; solid, the equilibrium
relation with its Maxwell plateaus.}
\label{fig:phases}
\end{figure*}

Figure~\ref{fig:phases}(a) gives, for each $D$, the free energy of the
lowest local minimum in each of the three basins.
The $\delta$ and $\beta$ basins stay stiff over the range shown
(Table~\ref{tab:barriers}), so each minimizing configuration is the
relaxed structure of that phase and each curve is the parabola
\begin{equation}
\Phi_X(D)=U_X+K_X\,(D-P_X)^2,\qquad
K_X=\frac{\Omega_X}{2\varepsilon_0\varepsilon_\infty^{X}},
\label{eq:parabola}
\end{equation}
with $U_X$, $P_X$, $\Omega_X$, and $\varepsilon_\infty^X$ taken from
Table~\ref{tab:phases}.
The same form is used for $\alpha$ at $D\le D_{c1}$.
The curvatures are $38.0$, $37.5$, and
$35.3$~eV\,m$^4$\,C$^{-2}$ for $\alpha$, $\delta$, and $\beta$, and
agree to $7\%$.

The global minimum passes from $\alpha$ to $\delta$ at
\begin{equation}
D_{c1}=0.046~\text{\Cm},
\end{equation}
and from $\delta$ to $\beta$ at
\begin{equation}
D_{c2}=0.157~\text{\Cm}.
\end{equation}
For equal $K$ the crossings are
$D_c=\frac12(P_1+P_2)+s$, with
$s\equiv\varepsilon_0\varepsilon_\infty\Delta U/\Omega\Delta P$.
The offsets are $-0.002$~\Cm\ for $\alpha$--$\delta$ and
$+0.013$~\Cm\ for $\delta$--$\beta$.
$D_{c1}$ and $D_{c2}$ are therefore set by the polarizations, and only
weakly by the internal-energy differences that carry the main $E=0$
uncertainty.

Thus $D$ alone---one thermodynamic variable, even though the two
steps run along different coordinates---orders the phases as
$\alpha\to\delta\to\beta$.
The first step is the conversion seen on poling $\alpha$ film
\cite{davis1978,li2013}.
The second is not: $\beta$ is reached in practice by drawing, or from
$\alpha$ by field alone only at substantially higher $E$
(Sec.~\ref{sec:intro}).
The ordering names the global minimum at each $D$; it does not say
whether the crystal can reach that minimum, and there the two steps
differ sharply (Sec.~\ref{sec:barriers}).

The crossings of the homogeneous parabolas are not the equilibrium
construction.
$\Phi_{\min}(D)$ is not a convex function of $D$: a straight line
joining two of its points can lie below the curve, as the line
joining the two $\delta$ minima does at $D=0$.
When the free energy fails to be convex in $D$, the system lowers
$\Phi$ by separating into two phases, and the equilibrium free energy
is the convex hull
\cite{stengel2009,stengel2009b,hong2011}.

The hull is the correct construction for one interface geometry only.
Suppose the phases form lamellae whose interfaces are parallel
to $c$, so that $E$ and $D$, both directed along the polar axis, lie
in the interface plane.
The tangential component of $E$ is continuous across an interface, so
the two phases share a single field; $D$ is under no such constraint
and may differ between them.
Both the displacement and the free energy are then volume-weighted
averages over the two phases, so a mixture traces a straight line
between their two points in $\Phi(D)$---the line the hull
construction assumes.
Had the interfaces been perpendicular to $c$, the roles would be
exchanged: $D$ would be the continuous variable and common to both
phases, $E$ would differ between them, and two phases at different
displacements would be impossible.
That problem is capacitors in series, not phase coexistence.
Throughout this section, coexistence therefore means lamellae (or
domains) running parallel to $c$.

On that geometry the convex hull of $\Phi_{\min}(D)$, extended to
$-D$ by symmetry, is the free energy when the crystal may break into
domains.
Its slope is the coexistence field
$E^*=\Omega^{-1}\,d\Phi/dD$, the common electric field of the two
phases on the tie line.
Wherever the hull lies below the homogeneous curve, the equilibrium
state is a two-phase mixture at that constant $E^*$ (Maxwell
construction).
A subscript on $E^*$ names the pair of phases when more than one tie
line is in play.

Two nonconvex regions appear in Fig.~\ref{fig:phases}(a).
(i) For $|D|<P_\delta=0.096$~\Cm\ the hull is the horizontal line $\Phi=U_\delta$ joining the $\pm P_\delta$ minima of $\delta$.
(ii) For $0.11<D<0.21$~\Cm\ the hull is the common tangent between the $\delta$ and $\beta$ minima.
We take them in turn.

(i) For $|D|<P_\delta=0.096$~\Cm\ the hull is the horizontal line
$\Phi=U_\delta$ joining the $\pm P_\delta$ minima of $\delta$.
That is a crystal split into antiparallel $\delta$ domains: a tie
line of zero slope, $E^*=0$, and $s=0$.
Each domain sits at $D=\pm P_\delta$, where $E=0$, so the mixture
pays no depolarization energy and costs only $U_\delta=-14.3$~meV per
cell, below homogeneous $\alpha$ at $\Phi=0$.
Figure~\ref{fig:landscape}(c) shows the homogeneous contrast.
Every curve there is a single-domain state, and at $D=0$ homogeneous
$\delta$ lies $334$~meV above $\alpha$ because it carries the
full $K_\delta P_\delta^2$.
The domain state avoids that penalty and is the state on the hull.

Wall energy is neglected, as in every construction of this kind.
That is the large-domain limit: a wall costs an energy proportional
to its area, so the cost per volume falls as the inverse domain size.
At finite size it raises the hull, most at equal domain fractions and
not at all at the ends of the tie line.
In PVDF the wall is the conformational kink that carries the
boundary \cite{dveyaharon1980}.
That is one more reason not to press the $14.3$~meV margin.

On these terms---this functional, zero temperature, no wall
energy---a $\delta$ multidomain state lies $14.3$~meV per cell below
homogeneous $\alpha$ at every $D$.
Homogeneous $\alpha$ is never the ground state.
The margin is small, of the same order as the functional spread of
$U_\delta-U_\alpha$ \cite{pelizza2016,kim2016,geng2025} and as the
vibrational contribution.
A $\Gamma$-point estimate (Sec.~\ref{sec:discussion}) suggests that
the ordering reverses near $220$~K, leaving $\alpha$ lower at room
temperature.
The conclusion is not that $\alpha$ is metastable in the laboratory,
but that melt-grown $\alpha$ is not explained by the
zero-temperature energetics.
It is a matter of crystallization kinetics and of the barrier in
Sec.~\ref{sec:barriers}.
If the ordering reverses, a sliver of $\alpha+\delta$ coexistence
with a small finite $E^*$ replaces the flat tie line; the rest of
the construction is unchanged.

(ii) For $0.11<D<0.21$~\Cm\ the hull is the common tangent between
the $\delta$ and $\beta$ minima.
For the parabolas of Eq.~(\ref{eq:parabola}) that tangent can be
written in closed form.
Equal slopes at the two contact points require $D_1=P_\delta+s$ and
$D_2=P_\beta+s$ with the same offset $s$ as above, and therefore
\begin{align}
D_2-D_1&=\Delta P, \qquad \tfrac12(D_1+D_2)=D_{c2},\notag\\
E^*_{\delta\beta}&=\frac{\Delta U}{\Omega\,\Delta P}=0.69~\text{GV\,m}^{-1}.
\label{eq:tie}
\end{align}
The window is as wide as the polarization difference it must supply,
$\Delta P=0.098$~\Cm, and is centered on $D_{c2}$.
In each phase the field is $(D_i-P_i)/\varepsilon_0\varepsilon_\infty
=s/\varepsilon_0\varepsilon_\infty=E^*$, the common field required for
lamellae parallel to $c$.
Two properties of $E^*$ follow from Eq.~(\ref{eq:tie}).

First, $E^*$ does not depend on the dielectric constant.
It is fixed by $\Delta U$ and $\Delta P$ alone and is therefore the
one quantity in this section that is insensitive to the clamped-ion
approximation, whereas $D_{c2}$ shifts with the screening through $s$
(Sec.~\ref{sec:screening}).

Second, $E^*$ is the same field one would write at fixed $E$ without
a hull.
The electric enthalpy of a stiff phase is $U_X-\Omega E P_X$, and the
two enthalpies are equal at $E=\Delta U/\Omega\Delta P$, which is
Eq.~(\ref{eq:tie}).
That is a consistency check: the common-tangent conditions are
equality of $E$ and of the enthalpy density $\Phi/\Omega-ED$, and the
second of these is the fixed-$E$ crossing.
In a fixed-$D$ capacitor the $\delta\to\beta$ conversion therefore
runs through a coexistence window in which the fraction of $\beta$
grows linearly with $D$ at constant $E=E^*$.

This $E=E^*$ segment is the plateau of a first-order field-induced
transition between two polar phases.
The common-tangent construction is the same as for an
antiferroelectric antipolar-to-polar transition
\cite{kittel1951,toledano2016}, but the loop is not pinched at the
origin: the plateau is entered from polar $\delta$, not from a
zero-polarization state.
The step is not reversible.
The reverse barrier is still $0.36$~eV per cell at $D=0$
(Table~\ref{tab:barriers}), so $\beta$ is retained when the field is
removed, as in drawn $\beta$ film.

Figure~\ref{fig:phases}(b) is the corresponding $E(D)$.
The homogeneous branches are linear with slope
$1/\varepsilon_0\varepsilon_\infty$ and jump at each homogeneous
crossing.
The equilibrium curve replaces those jumps by the Maxwell plateaus.
Fields of several GV\,m$^{-1}$ on the homogeneous branches belong to
the clamped-ion, single-coordinate model
(Sec.~\ref{sec:screening}) and are not coercive fields.

$E^*_{\delta\beta}$ should also not be compared with the
$\simeq 0.5$~GV\,m$^{-1}$ at which $\beta$ is obtained from $\alpha$
films \cite{davis1978,geng2025}.
Those films are poled from metastable $\alpha$, so the relevant tie
line is $E^*_{\alpha\beta}=(U_\beta-U_\alpha)/\Omega P_\beta
=0.29$~GV\,m$^{-1}$ (Sec.~\ref{sec:phases}), a factor of $1.7$ below
the observed field.
That is the expected sense: a thermodynamic field is a lower bound on
a measured coercive field, and the difference is the kinetic overhead
of nucleation and barrier crossing (Sec.~\ref{sec:barriers}).
$E^*_{\delta\beta}$ is a different quantity, the plateau field of a
sample already in $\delta$.
It has no direct experimental counterpart, because $\beta$ film is
drawn rather than converted from $\delta$ by field alone.

The $\delta+\beta$ window uses two further assumptions.
Interfacial energy is neglected on the same large-domain grounds as
in (i).
Elastic energy is not of that kind.
The cells differ by $-9.5\%$ in $a$ and $+10.6\%$ in $b$, so coherency
strain scales with volume, not with interface area, and does not
vanish for large domains.
Coherent coexistence at that mismatch would be prohibitive.
The construction therefore applies to a semicrystalline film, in
which the lever-rule lamellae can be separated by amorphous material
and need not be coherent, rather than to a single crystal.
The $\pm\delta$ construction of (i) has no such elastic caveat: the
two domains are the same phase in the same cell.

\subsection{Robustness to lattice screening}
\label{sec:screening}

Equation~(\ref{eq:phi}) screens the depolarizing field with the
clamped-ion $\varepsilon_\infty$ only.
That is adequate when the coordinate is itself the polar soft mode,
because the main ionic response then lies in $\lambda$.
Here the tilt and torsion are not eigenmodes, and polar modes
orthogonal to them---chain librations and the lowest optic modes at
$14$--$40$~cm$^{-1}$---also respond to $E$.

The replacement for $\varepsilon_\infty$ is not the measured film
constant ($\simeq 10$ at $1$~kHz \cite{kepler1992}).
That value includes the reorientation of the chain dipoles, which
$\Phi$ already treats through $\lambda$, and the relaxations of the
amorphous fraction.
Using it would count the path twice.
The appropriate constant is
$\varepsilon_{\rm eff}=\varepsilon_\infty+\Delta\varepsilon_\perp$,
where $\Delta\varepsilon_\perp$ is the lattice contribution of the
modes orthogonal to $\lambda$ only.

That quantity is computable.
Let $\delta\bm{u}$ be the infinitesimal atomic displacement induced
by a field $E$, $\mathsf{C}$ the zone-center force-constant matrix
with rigid translations projected out, and $\tilde Z$ the $3N$-vector
of Born charges $Z^*_{\kappa,cj}$ that give $P_c$ per unit
$\delta\bm{u}$.
Let $\hat{\bm e}$ be the unit tangent of the path (internal
displacement between neighboring scan images, strain removed,
normalized).
The lattice response at fixed $\lambda$ is the constrained
minimization of
$\tfrac12\delta\bm{u}^{\rm T}\mathsf{C}\,\delta\bm{u}
-\tilde Z^{\rm T}\delta\bm{u}\,E$
subject to $\hat{\bm e}\cdot\delta\bm{u}=0$.
Eliminating the Lagrange multiplier gives, in atomic units,
\begin{equation}
\Delta\varepsilon_\perp=\frac{4\pi}{\Omega}\left[
\tilde Z^{\rm T}\mathsf{C}^{-1}\tilde Z
-\frac{(\tilde Z^{\rm T}\mathsf{C}^{-1}\hat{\bm e})^2}
{\hat{\bm e}^{\rm T}\mathsf{C}^{-1}\hat{\bm e}}\right].
\label{eq:epsperp}
\end{equation}
The first term is the unconstrained lattice response
$\Delta\varepsilon_{\rm latt}$.
The second subtracts the piece the field would have obtained by
moving along $\hat{\bm e}$---the double counting already in
$\Phi(\lambda)$.
Table~\ref{tab:eps} lists the result at the three relaxed
structures.

The polar-axis lattice response is small.
The full clamped-cell static constant
$\varepsilon_\infty+\Delta\varepsilon_{\rm latt}$ is only $2.4$--$2.8$,
and the part not already in the path raises the screening constant
by $12$--$26\%$.
The response is smallest in $\beta$ because it is carried by chain
libration: in $\beta$ the chain dipole already lies along $c$, so
libration changes $P_c$ only to second order
($\Delta\varepsilon_{\rm latt}=1.39$ along $a$ and $0.27$ along
$c$).
The torsion tangent at $\beta$ has no overlap with the
field-induced displacement, and
$\Delta\varepsilon_\perp=\Delta\varepsilon_{\rm latt}$ there.
These are clamped-cell values at the relaxed phases.
A free-stress constant would add a piezoelectric term, and the
softer saddle configurations were not treated by perturbation
theory.
Neither limitation should change the order of a correction that is
itself a quarter of $\varepsilon_\infty$.

The consequences for the results above are correspondingly modest.
We test them with two screening constants in place of the clamped-ion
$\varepsilon_\infty$ used throughout.
The first is the computed $\varepsilon_{\rm eff}=\varepsilon_\infty+
\Delta\varepsilon_\perp$, about $1.25\,\varepsilon_\infty$
(Table~\ref{tab:eps}), which is our best estimate of the screening.
The second is $2\varepsilon_\infty$, a deliberately extreme value,
$1.6$ times the largest $\varepsilon_{\rm eff}$.
Table~\ref{tab:barriers} gives the results for both, and the dotted
lines in Fig.~\ref{fig:barriers} show those for $2\varepsilon_\infty$.

Three classes of quantity respond differently.
The coexistence fields $E^*$ do not depend on the dielectric constant
[Eq.~(\ref{eq:tie})].
With $\varepsilon_{\rm eff}$ the closed-form $E^*_{\delta\beta}$ is
unchanged at $0.689$~GV\,m$^{-1}$, and even $2\varepsilon_\infty$
changes the numerical-hull value only from $0.686$ to
$0.684$~GV\,m$^{-1}$.
The critical displacements shift only through the offset
$s=\varepsilon_0\varepsilon\Delta U/\Omega\Delta P$, which is linear in
$\varepsilon$.
With $\varepsilon_{\rm eff}$, $D_{c1}$ stays at $0.046$~\Cm\ and
$D_{c2}$ moves from $0.157$ to $0.161$~\Cm.
With $2\varepsilon_\infty$, $D_{c2}$ moves to $0.170$~\Cm.
The fields of single-domain states, $E=(D-P_X)/\varepsilon_0\varepsilon$
(the homogeneous branches of Fig.~\ref{fig:phases}(b)), scale as
$1/\varepsilon$ and are the most affected of the three.

The barriers rise with screening, but their contrast survives.
With $\varepsilon_{\rm eff}$, both barriers at $D=0.30$~\Cm\ rise
slightly.
The tilt barrier goes from $0.02$ to $0.06$~eV, and the torsion
barrier from $0.34$ to $0.40$~eV.
The tilt barrier still drops below room-temperature $k_BT$, now at
$D=0.35$ instead of $0.28$~\Cm.
With $2\varepsilon_\infty$, the two barriers at $D=0.30$~\Cm\ are
$0.21$ and $0.51$~eV.
The tilt barrier is then no longer removed within the range, but it
stays far below the torsion barrier.

Screening also lowers the depolarization penalty of $\beta$, but only
modestly.
At $D=0$, $\Udp$ of $\beta$ is $1.33$~eV with $\varepsilon_\infty$.
With $\beta$'s own $\varepsilon_{\rm eff}=2.42$ it is $1.18$~eV, still
$15$ times the $\beta$--$\alpha$ internal-energy gap and $1.5$ times
the $E=0$ torsion barrier.
In every case the trends are moderated, not reversed.

No homogeneous polar phase is the $D=0$ ground state at the
clamped-ion $\varepsilon_\infty$, and none becomes so under any
plausible increase in screening.
For single-domain $\delta$ to fall below $\alpha$ at $D=0$ the
constant would have to exceed $52$, far above $\varepsilon_{\rm eff}$
and above the film value that double-counts the chain reorientation.
For $\beta$ no constant suffices, because $\beta$ already lies above
$\alpha$ in internal energy.

\section{Discussion}
\label{sec:discussion}

\emph{Relation to processing.}
The results give a thermodynamic reading of the empirical route to
$\beta$-PVDF.
The $\alpha\to\delta$ step changes the cell by less than $1\%$ and
is driven by the coupling of $D$ to the chain dipoles
(Sec.~\ref{sec:ad0}).
$D$ can remove its barrier, so no intrinsic barrier on the coordinate
limits the conversion; at experimental fields the bottleneck is
nucleation (Sec.~\ref{sec:barriers}).
That is the poling step.

The $\delta\to\beta$ step is inseparable from a $10.6\%$ elongation
along the chain axis.
It remains first order at every $D$, and its barrier is still
$0.66$~eV per cell at $D_{c2}$.
A field must cross that barrier, either coherently or by nucleation.
Drawing supplies the strain that is the other half of the
coordinate.
In practice $\beta$ film is made in the reverse order.
Drawing $\alpha$ straightens the chains, and subsequent poling only
aligns the $\beta$ crystallites \cite{lovinger1983,kepler1992},
reversing chains by rotation about their axes (Sec.~\ref{sec:coords}).
Whichever comes first, a field drives the reversal and strain drives
the torsion.

Section~\ref{sec:zstar} identifies the electrical side of this
division.
Per unit atomic displacement, the tilt changes the polarization about
$1.5$ times as much as the torsion.
That larger effective charge is what allows $D$ to remove the tilt
barrier: with the effective charge of the torsion, the tilt would keep
a barrier of five times $k_BT$ at $D=0.30$~\Cm.
The torsion has a smaller effective charge, a higher starting barrier,
and a dilated saddle, and $D$ lowers its barrier without removing it.
The calculation does not derive the processing route, but it accounts
for the split between poling and drawing.

\emph{Toward field-driven $\beta$.}
The same analysis points to a way of removing that split by chemistry.
In the model of Sec.~\ref{sec:zstar}, doubling the effective charge of
the torsion at fixed internal energy and cell lowers its barrier at
$D=0.30$~\Cm\ from $0.34$ to $0.07$~eV, and tripling it removes the
barrier.
A field would then carry $\delta$ into $\beta$ without drawing, in the
same coherent sense in which it carries $\alpha$ into $\delta$.
Doubling the effective charge also halves the coexistence field
$E^*_{\delta\beta}=\Delta U/\Omega\Delta P$ of Eq.~(\ref{eq:tie}), from
$0.69$ to $0.34$~GV\,m$^{-1}$, because $\Delta P$ doubles while
$\Delta U$ does not.
A substitution that raises the polarization change of the
conformational step without adding to its atomic displacement is
therefore one route to a polymer whose polar all-trans phase is
reached by a field alone.

Two cautions apply.
First, the scaling holds $\UKS$, $\Omega$, and $\varepsilon_\infty$
fixed.
A real substitution changes them as well, in particular $\Delta U$
and the chain-axis strain, which carry the other half of the
response of the torsion.
Second, a larger $P_\beta$ costs stability at open circuit.
For $\eta\ge1.75$, $\beta$ is not a local minimum at $D=0$, so a
$\beta$ phase made in this way would be retained under short circuit,
where its reverse barrier is the $0.71$~eV of $\UKS$, but not at open
circuit.
Comonomers in current use, such as trifluoroethylene, favor the
all-trans conformation mainly by lowering $\Delta U$
\cite{lovinger1983,furukawa1989}.
The effective charge is a separate lever and can be combined with
that one.
Because $Z^*_{\rm path}$ requires only the two end points and their
branch-resolved polarizations, it can be used to screen candidate
chemistries before any barrier is computed, together with the $E=0$
barrier and strain that set the rest of the response.

\emph{Antiferroelectric character of the first step.}
The $\alpha$ phase is an antipolar packing of polar chains, and
$\alpha\to\delta$ under a field is formally an
antiferroelectric-to-ferroelectric conversion in Kittel's sense
\cite{kittel1951,toledano2016}: two sublattices (the two chains)
carry equal and opposite dipoles that the field aligns.
In a conventional antiferroelectric the antipolar phase is the ground
state, so a finite field is needed to reach the polar phase, which
gives the double hysteresis loop.
PVDF differs in two respects.
The two phases share a cell and are related by the setting of one
chain.
They also differ in energy by only $3.6$~meV per monomer, and at $T=0$
the polar $\delta$ is the lower.
The $\pm\delta$ domain state therefore lies below homogeneous $\alpha$
at every $|D|<P_\delta$, and there is no $\alpha+\delta$ window with a
finite $E^*$ (Sec.~\ref{sec:sequence}).
Because the margin is so small, a window can open at finite
temperature.
The $\Gamma$-point estimate of Limitation~(4) suggests that $\alpha$
becomes the lower phase near $220$~K.
At room temperature the conversion is then set by that narrow window,
by the tilt barrier, and by domain processes.

\emph{Energy storage.}
The area under the equilibrium $E(D)$ of Fig.~\ref{fig:phases}(b)
is the electrostatic energy stored per unit volume.
Charging an ideal crystal to the end of the $\delta+\beta$
coexistence window ($D=0.21$~\Cm) stores $72$~J\,cm$^{-3}$.
Almost all of it, $E^*\Delta P=67$~J\,cm$^{-3}$, is the
$\delta\leftrightarrow\beta$ plateau.
By Eq.~(\ref{eq:tie}) that plateau term equals $\Delta U/\Omega$,
the internal-energy cost of the conformational change, not the
dielectric response of either phase.
The remaining $5$~J\,cm$^{-3}$ is dielectric charging of
homogeneous $\delta$ from $D=P_\delta$ to the onset of coexistence
$D_1=P_\delta+s$, and is the only piece that depends on the screening
constant.
The $\pm\delta$ domain state below $|D|=P_\delta$ is traversed at
$E^*=0$ and stores nothing in equilibrium.

That $72$~J\,cm$^{-3}$ is an upper bound for an ideal, fully
crystalline, oriented sample switched reversibly.
It requires $E^*=0.69$~GV\,m$^{-1}$, at or above the breakdown of
ordinary PVDF films and exceeded only in the exceptional films of
Ref.~\cite{meng2019}.
Recovering the energy in a real cycle means recrossing the
$0.36$~eV reverse barrier of Table~\ref{tab:barriers}.
For comparison, computed values are $\sim 20$~J\,cm$^{-3}$ at
$0.5$~GV\,m$^{-1}$ for disordered P(VDF--CTFE) \cite{ranjan2007}
and $59$~J\,cm$^{-3}$ at $0.45$~GV\,m$^{-1}$ for a VDF--VF
copolymer \cite{geng2025}.
The highest measured discharged energy of the homopolymer
is $35$~J\,cm$^{-3}$ at $0.88$~GV\,m$^{-1}$
\cite{meng2019}, against $17$~J\,cm$^{-3}$ at $0.575$~GV\,m$^{-1}$
for P(VDF--CTFE) \cite{chu2006} and $\simeq 1$--$2$~J\,cm$^{-3}$
for biaxially oriented polypropylene.

To bring the plateau inside the breakdown limit, chemistry must lower
$E^*_{\delta\beta}=\Delta U/\Omega\Delta P$.
A comonomer that reduces $\Delta U$ does so
\cite{ranjan2012,dong2016,geng2025}, but it lowers the plateau energy
$\Delta U/\Omega$ in the same proportion.
A larger $\Delta P$, the effective-charge lever above, lowers $E^*$
and leaves the plateau energy unchanged.

\emph{Limitations.}
(1) All barriers are upper bounds, because the carbon skeleton is
confined to a chosen coordinate and the 24-atom cell forces a
concerted transformation (Sec.~\ref{sec:coords}).
Lower-barrier routes are localized.
For $\alpha\to\delta$ they include a propagating twist
\cite{dveyaharon1980} and the nonconcerted path of
Ref.~\cite{kim2016}, whose barrier is lower by a factor of $3$--$7$
(Sec.~\ref{sec:ad0}).
For $\delta\to\beta$ a sequential gauche flip could pass through
$\gamma$-like T$_3$G intermediates in a doubled cell.
Both require a long supercell along the chain.
$D_{c1}$, $D_{c2}$, and the coexistence fields $E^*$ do not depend on
the barriers and are unaffected.

(2) The screening in Eq.~(\ref{eq:phi}) is clamped-ion.
The lattice response orthogonal to the path adds $12$--$26\%$ at the
relaxed phases (Sec.~\ref{sec:screening}).
Strain response and the softer saddle modes remain untreated; of the
results, only the homogeneous-branch fields $E(D)$ are sensitive to
those omissions.

(3) The $E=0$ energetics carry the uncertainty of the functional.
$E^*_{\delta\beta}=\Delta U_{\beta\delta}/\Omega\Delta P$ inherits
that uncertainty linearly.
$D_{c2}$ shifts only through the offset $s\propto\Delta U$, which is
under 10\% of $D_{c2}$.

(4) All energies above are internal energies at $T=0$.
At finite $T$ the well depth $U_X$ in Eq.~(\ref{eq:parabola}) can be
replaced by the vibrational Helmholtz free energy, $F_X(T)=U_X+F^X_{\rm vib}(T)$.
In the harmonic approximation
\begin{equation}
F_{\rm vib}(T)=\sum_\nu\left[\frac{\hbar\omega_\nu}{2}
+k_BT\ln\!\left(1-e^{-\hbar\omega_\nu/k_BT}\right)\right].
\label{eq:fvib}
\end{equation}
The first term is the zero-point energy; the second is the thermal
occupation of the modes $\omega_\nu$.
$E^*$ and the critical displacements depend on the phase energies
only through differences, so Sec.~\ref{sec:sequence} carries over
with $\Delta U\to\Delta F(T)$.
Thermal expansion and the $T$ dependence of $P_X$ are not treated.

This matters most for the $\alpha$/$\delta$ ordering, whose margin
is $14.3$~meV per cell ($3.6$~meV per monomer).
Evaluating Eq.~(\ref{eq:fvib}) on the $69$ zone-center optic modes
(three acoustic modes removed) gives a zero-point energy $3.7$~meV
per cell higher in $\delta$, whose two lowest optic modes ($40$ and
$59$~cm$^{-1}$) are stiffer than those of $\alpha$ ($33$ and
$51$~cm$^{-1}$).
$F_\delta-F_\alpha$ is then $-10.6$~meV per cell at $T=0$, crosses
zero near $220$~K, and is $+5$~meV per cell at $300$~K, with ten
soft modes accounting for $97\%$ of the difference.
A $\Gamma$-point estimate on a $24$-atom cell samples the
low-frequency branches poorly, so it fixes the sign of the effect
and not the exact crossing temperature.
It still makes a room-temperature reversal of the $T=0$
$\alpha$/$\delta$ ordering in Table~\ref{tab:phases} likely.

(5) $\Phi$ is evaluated on $E=0$ geometries, so only the scanned 
coordinate (the tilt or the torsion) responds to $D$.
That is accurate in the stiff $\delta$ and $\beta$ wells and is 
an estimate at the saddles.
A constrained-$D$ relaxation \cite{stengel2009} would lift the 
restriction but would lose the phase-resolved landscape.

\section{Conclusions}
\label{sec:conclusions}

We have determined the free energy of PVDF under finite electric
displacement along two collective coordinates that share the relaxed
$\delta$ phase: a tilt that converts $\alpha$ into $\delta$, and a
torsion that converts $\delta$ into $\beta$.
The calculations use a van der Waals functional and a Berry-phase
polarization whose branch is tracked along the path.
The findings are as follows.

(i) The polarization of $\beta$ is $0.55$ of the quantum $e/ab$ of
the 24-atom cell, so the raw Berry phases of isolated configurations
cannot be read at face value.
The branch is tracked along the physical path and anchored at
centrosymmetric $\alpha$, where $P=0$ by inversion, so no artificial
reference structure is needed.

(ii) At $E=0$ the two steps differ in kind.
The tilt reorients one chain, changes the cell by less than $1\%$,
and has a barrier of $0.58$~eV per cell.
It removes the cancellation of the dipoles between chains.
The torsion converts both chains from TG$^+$TG$^-$ to TTTT,
lengthens the chain repeat by $10.6\%$, and has a barrier of
$0.80$~eV.
It completes the alignment of the dipoles within each chain.

(iii) $D$ reorders the wells much more than it lowers the saddles
between them.
The depolarization energy is large because $\varepsilon_\infty\simeq2.2$
is small: at $D=0$ it is $1.33$~eV per cell for $\beta$.
Each saddle, however, has a polarization close to its own critical
displacement, so near that $D$ it pays almost no depolarization energy
and $\UKS$ supplies the barrier.
The tilt barrier falls from $0.66$~eV at $D=0$ to $17$~meV at
$D=0.30$~\Cm, below $k_BT$ at $300$~K.
The torsion barrier is still $0.66$~eV at $D_{c2}$ and $0.34$~eV at
$D=0.30$~\Cm, so $\delta\to\beta$ remains first order and hysteretic.
Once formed, $\beta$ is retained at $D=0$ behind a reverse barrier of
$0.36$~eV.

(iv) The path effective charge, the dipole created per unit atomic
displacement, accounts for much of this contrast.
It is $0.28e$ for the tilt and $0.18e$ for the torsion.
Scaling the polarization change along each path at fixed internal
energy separates its role.
With the effective charge of the torsion, the tilt keeps a barrier of
$0.13$~eV at $D=0.30$~\Cm.
With the effective charge of the tilt, the torsion keeps $0.16$~eV;
its higher starting barrier and dilated saddle account for the rest.
A torsion with three times its calculated effective charge would be
flattened by $D=0.30$~\Cm, but $\beta$ would then not be metastable
at open circuit.

(v) $D$ alone orders the phases $\alpha\to\delta\to\beta$, with
$D_{c1}=0.046$~\Cm\ and $D_{c2}=0.157$~\Cm.
The crossings are set by the polarizations; the internal-energy
differences enter only as offsets of $-0.002$ and $+0.013$~\Cm.
At $T=0$, with domain-wall energy neglected, a $\delta$ multidomain
state lies $14.3$~meV per cell below homogeneous $\alpha$ at every
$D$.
A $\Gamma$-point harmonic estimate of the vibrational free energy is
likely to reverse that ordering by room temperature.

(vi) $\delta$ and $\beta$ coexist for $0.11<D<0.21$~\Cm\ at
$E^*_{\delta\beta}=0.69$~GV\,m$^{-1}$.
This field is $\Delta U/\Omega\Delta P$ and does not depend on the
screening constant.
The corresponding $\alpha\to\beta$ field, $0.29$~GV\,m$^{-1}$, lies
below the $\simeq0.5$~GV\,m$^{-1}$ at which $\beta$ appears in poled
$\alpha$ films, as a thermodynamic bound should.

(vii) The zone-center lattice response orthogonal to the path raises
the screening constant by $12$--$26\%$ at the relaxed phases.
That correction, and even a doubling of $\varepsilon_\infty$, leaves
$E^*$, the phase sequence, and the first-order character of
$\delta\to\beta$ unchanged, and moves $D_{c2}$ by at most
$0.013$~\Cm.

These results account for the processing route: a field produces
$\delta$, and drawing produces $\beta$ by supplying the chain-axis
strain that is part of the torsion.
Cycling an ideal crystal through the $\delta\leftrightarrow\beta$
plateau stores at most $72$~J\,cm$^{-3}$, almost all of it the plateau
term $E^*\Delta P=\Delta U/\Omega$.
The path effective charge points to a design route toward polymers
whose polar all-trans phase is reached by a field alone: raise the
polarization change of the conformational step per unit atomic
displacement, for example by substitution.
In the same model this also lowers $E^*$ at constant stored energy.
The present barriers are concerted upper bounds, and supercells are
needed for the inhomogeneous mechanisms that set the observed
coercive fields.
The critical displacements and coexistence fields do not depend on the
barriers and are unaffected by that limitation.

\begin{acknowledgments}
The support and resources from the Center for High Performance Computing at
the University of Utah are gratefully acknowledged.
\end{acknowledgments}

\newpage

\begin{table*}[tb]
\caption{Relaxed structures and $E=0$ properties of the three
polymorphs in the common frame, compared with experiment
\cite{hasegawa1972} and a previous vdW-DF study \cite{pelizza2016}.
The experimental chain repeat of \bph\ is doubled ($2\times2.56$~\AA)
to match the cell.
$\Delta U$ is per cell of four monomers ($1$~meV per cell $=0.0241$~kJ\,mol$^{-1}$
per monomer).
$P_c$ is on the \aph-continuous branch (Sec.~\ref{sec:branch}).
$\varepsilon_\infty$ is the clamped-ion tensor and $\omega_{\min}$ the
lowest optical zone-center frequency; no phase has an imaginary
zone-center mode.}
\label{tab:phases}
\begin{ruledtabular}
\begin{tabular}{lcccccccc}
 & $a$ (\AA) & $b$ (\AA) & $c$ (\AA) & $\Omega$ (\AA$^3$) & $\Delta U$ (meV) & $|P_c|$ (\Cm) & $\varepsilon_\infty^{aa},\varepsilon_\infty^{bb},\varepsilon_\infty^{cc}$ & $\omega_{\min}$ (cm$^{-1}$) \\
\hline
\aph\ (P2$_1$/c)  & 9.630 & 4.710 & 4.990 & 226.31 & 0       & 0     & 2.103, 2.253, 2.099 & 32.5 \\
\quad expt.       & 9.64  & 4.62  & 4.96  & 220.9  &         &       &                     &      \\
\quad vdW-DF      & 9.57  & 4.69  & 4.99  & 224.0  &         &       &                     &      \\
\dph\ (Pna2$_1$)  & 9.552 & 4.713 & 4.994 & 224.80 & $-14.3$ & 0.096 & 2.111, 2.261, 2.113 & 40.1 \\
\quad expt.       &       &       &       &        &         & 0.07\footnotemark[1] & &      \\
\quad vdW-DF      & 9.53  & 4.70  & 4.98  & 223.1  & $-12$   &       &                     &      \\
\bph\ (Amm2)      & 8.642 & 5.213 & 4.793 & 215.94 & $+78.3$ & 0.194 & 2.155, 2.370, 2.156 & 13.7 \\
\quad expt.       & 8.58  & 5.12  & 4.91  & 215.7  &         & 0.076--0.13\footnotemark[1] & & \\
\quad vdW-DF      & 8.62  & 5.20  & 4.80  & 215.2  & $+158$  & 0.198 &                     &      \\
\end{tabular}
\end{ruledtabular}
\footnotetext[1]{Measured remanent polarizations of poled semicrystalline films (\dph: Ref.~\cite{li2013}; \bph: Refs.~\cite{kepler1992,furukawa1989}) and, for \bph, oligomer crystals; see text.}
\end{table*}

\begin{table}[tb]
\caption{Fixed-$D$ barriers (eV per cell) along the two coordinates:
the highest $\Phi$ between two basin minima, relative to the starting
minimum.
The last two rows use the computed lattice screening
(Sec.~\ref{sec:screening}) and a doubled $\varepsilon_\infty$.}
\label{tab:barriers}
\begin{ruledtabular}
\begin{tabular}{lcccc}
$D$ (\Cm) & \aph$\to$\dph & \dph$\to$\aph & \dph$\to$\bph & \bph$\to$\dph \\
\hline
0      & 0.66 & 0.32 & 1.43 & 0.36 \\
0.046 ($D_{c1}$) & 0.49 & 0.50 & 1.16 & 0.40 \\
0.10   & 0.31 & 0.71 & 0.91 & 0.51 \\
0.157 ($D_{c2}$) & 0.21 & 0.99 & 0.66 & 0.66 \\
0.20   & 0.13 & 1.22 & 0.51 & 0.82 \\
0.30   & 0.017 & 1.79 & 0.34 & 1.32 \\
0.30 ($\varepsilon_{\rm eff}=1.25\,\varepsilon_\infty$) & 0.060 & 1.50 & 0.40 & 1.19 \\
0.30 ($2\varepsilon_\infty$) & 0.21 & 1.13 & 0.51 & 0.99 \\
\end{tabular}
\end{ruledtabular}
\end{table}

\begin{table}[tb]
\caption{Lattice screening along the polar axis at the relaxed phases.
$\Delta\varepsilon_{\rm latt}$ is the clamped-cell lattice contribution
to $\varepsilon^{cc}$, and $\Delta\varepsilon_\perp$ its part orthogonal
to the tangent of the indicated scan [Eq.~(\ref{eq:epsperp})].
$\varepsilon_{\rm eff}=\varepsilon_\infty+\Delta\varepsilon_\perp$
replaces $\varepsilon_\infty$ in Eq.~(\ref{eq:udp}).}
\label{tab:eps}
\begin{ruledtabular}
\begin{tabular}{llccccc}
phase & coordinate & $\varepsilon_\infty$ & $\Delta\varepsilon_{\rm latt}$ & $\Delta\varepsilon_\perp$ & $\varepsilon_{\rm eff}$ & $\varepsilon_{\rm eff}/\varepsilon_\infty$\\
\hline
\aph\ & tilt    & 2.099 & 0.66 & 0.51 & 2.61 & 1.24\\
\dph\ & tilt    & 2.113 & 0.68 & 0.55 & 2.66 & 1.26\\
\dph\ & torsion & 2.113 & 0.68 & 0.53 & 2.64 & 1.25\\
\bph\ & torsion & 2.156 & 0.27 & 0.27 & 2.42 & 1.12\\
\end{tabular}
\end{ruledtabular}
\end{table}

\end{document}